\documentclass[10pt,journal,compsoc]{IEEEtran}

\usepackage{booktabs}

\ifCLASSOPTIONcompsoc
  \usepackage[nocompress]{cite}
\else
  \usepackage{cite}
\fi

\usepackage{multirow}
\usepackage{xcolor}
\usepackage{soul}
\usepackage{url}
\usepackage{subcaption}
\usepackage[section]{placeins}

\usepackage{longtable}

\usepackage[colorinlistoftodos,prependcaption,textsize=tiny]{todonotes}

\ifCLASSINFOpdf
\else
\fi

\begin{document}
%
\title{Impact of annotation modality on label quality and model performance in the automatic assessment of laughter in-the-wild}
%
%
%
%

\author{Jose~Vargas-Quiros,
        Laura~Cabrera-Quiros,
        Catharine~Oertel,
        and~Hayley~Hung
\IEEEcompsocitemizethanks{\IEEEcompsocthanksitem J. Vargas, C. Oertel and H. Hung work at the Department of Intelligent Systems at TU Delft, The Netherlands.\protect\\
E-mail: {j.d.vargasquiros, c.r.m.m.oertel,  h.hung}@tudelft.nl
\IEEEcompsocthanksitem L. Cabrera Quiros works at Escuela de Ingenieria Electronica at the Instituto Tecnologico de Costa Rica, Costa Rica.}
\thanks{Manuscript received April 19, 2005; revised August 26, 2015.}}

%
%


\IEEEtitleabstractindextext{%
\begin{abstract}
Laughter is considered one of the most overt signals related to enjoyment of an interaction. Although laughter is well-recognized as a multimodal phenomenon that is most commonly detected by sensing its vocal manifestations (ie. the sound of laughter), it is unclear how perception and annotation of laughter may differ when annotated from other modalities like video, where the body movements of laughter can be perceived. This is particularly relevant for studies of laughter in the wild, where audio recordings may not be available. In this paper we take a first step in this direction by asking if and how well laughter can be annotated when only audio, only video (containing full body movement information) or audiovisual modalities are available to annotators. We ask whether annotations of laughter are congruent across modalities, and compare the effect that labeling modality has on machine learning model performance. We compare annotations and models for laughter detection, intensity estimation, and segmentation, three tasks common in previous studies of laughter. Our analysis is in the context of a challenging in-the-wild conversational dataset with a variety of camera angles, noise conditions and voices. Our statistical analysis of more than 4000 annotations acquired from 48 annotators revealed that laughter could be annotated from video with high precision. Inter-annotator agreement revealed evidence for incongruity in the perception of laughter, and its intensity between modalities. Further analysis of annotations against consolidated audiovisual reference annotations revealed that recall was lower on average for video when compared to the audio condition, but tended to increase with the intensity of the laughter samples. Our machine learning experiments compared the performance of state-of-the-art unimodal (audio-based, video-based and acceleration-based) and multi-modal models for different combinations of input modalities, training label modality, and testing label modality, for a total of more than 120 model evaluations. Models with video and acceleration inputs had similar performance regardless of training label modality, suggesting that it may be entirely appropriate to train models for laughter detection from body movements using video-acquired labels, despite the lower inter-rater agreement measured in the video condition. For models with audio as input, training labels acquired in the audio and audiovisual condition resulted in the best overall scores.

\end{abstract}

}

\maketitle

\IEEEdisplaynontitleabstractindextext
\IEEEpeerreviewmaketitle

\IEEEraisesectionheading{\section{Introduction}\label{sec:introduction}}

Laughter is traditionally associated to its characteristic vocalization (ie. the sound of laughter). In research too, its vocal manifestation has received the most emphasis.

Nonetheless, laughter is a multimodal phenomenon. Darwin presented a curious depiction of excessive laughter: "the whole body is often thrown backwards and shakes, or is almost convulsed; the respiration is much disturbed; the head and face become gorged with blood; with the veins distended; and the orbicular muscles are spasmodically contracted in order to protect the eyes. Tears are freely shed." \cite[p.208]{Darwin1887}. This depiction makes reference to multiple characteristic manifestations of laughter: the facial movements of laughter, the full-body movements of laughter, and the physiological changes of laughter.


Following this premise, works in social signal processing \cite{Burgoon, Vinciarelli2009} have delved into the problem of automatically detecting and classifying laughter from audio, video and audiovisual recordings of its manifestations. Annotation is a key step in these studies. The first step in annotation of naturally occurring laughter usually involves the temporal localization, or segmentation of laughter (from its context). Next, laughter segments or episodes are categorized or otherwise rated. Functional or formal categorizations are the most common, but no consensus coding schemes exists for either of these tasks. Laughter intensity is also a common variable of interest that has been rated in multiple studies \cite{Mancini2012, Niewiadomski2012, McKeown2013, DiLascio2019, Niewiadomski2015, Haddad, Curran2018}. \cite{Mazzocconi2020} have linked laughter intensity directly to the meaning of laughter, as an indication of the magnitude of a positive shift in arousal caused by the laughable (the object of laughter) in the laughing subject.

Nevertheless, the emphasis on the vocal manifestations of laughter still translates strongly to its annotation, where laughter has most commonly been annotated from audio or audiovisual face recordings, by third-party observers \cite{Petridis2008a, Petridis2013, Truong2007, Truong2012}.

\begin{figure}
  \includegraphics[width=\columnwidth]{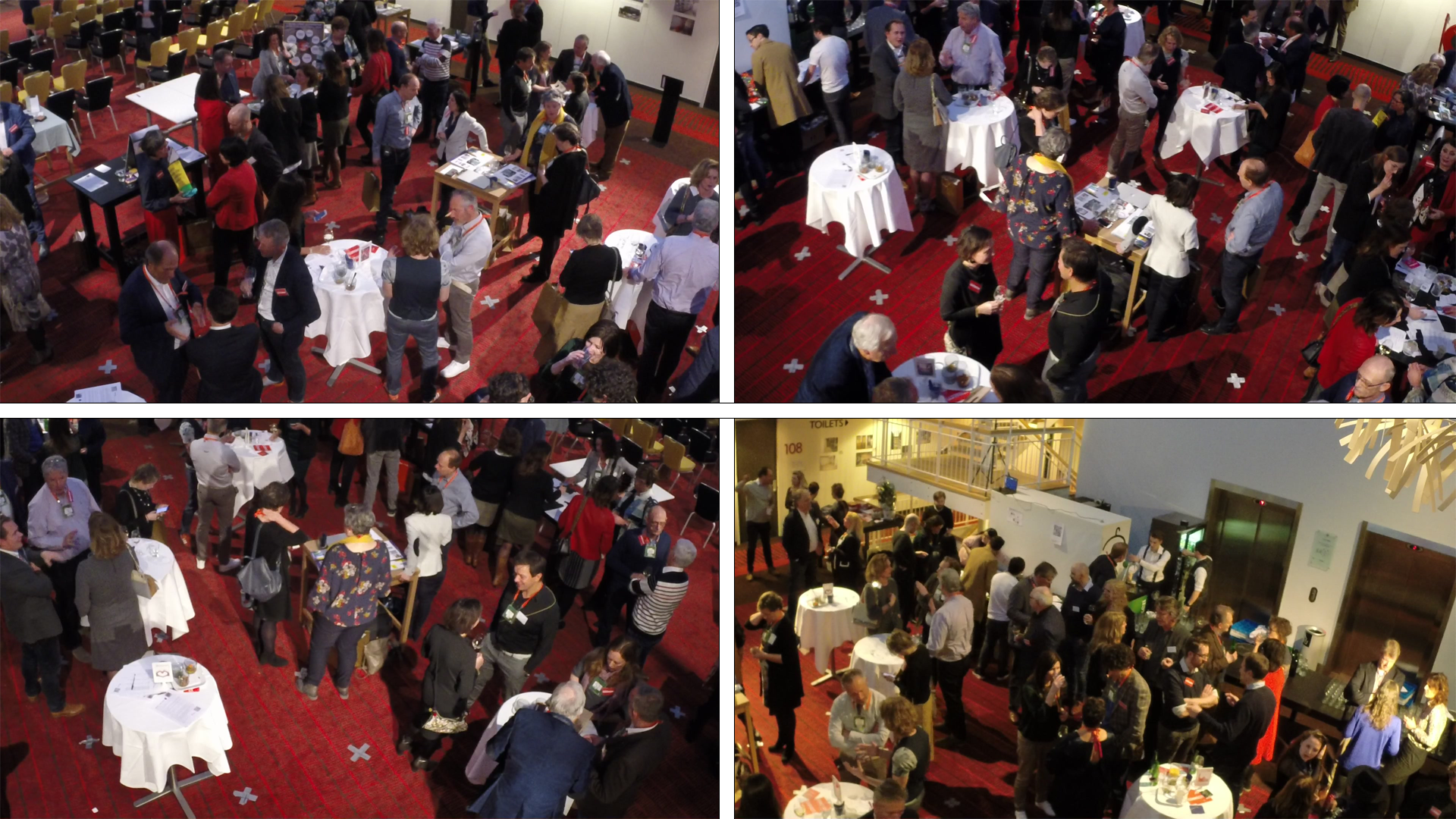}
  \caption{Screenshots from the four elevated views in our dataset of free-standing interactions, used for our human laughter annotation and automatic laughter assessment tasks.}
  \label{fig:lared}
\end{figure}

Laughter has also been annotated from body movements alone, using video. This has been done in in-the-wild datasets of \textit{mingling crowds} recorded in real-life events \cite{Cabrera-Quiros2018a}, such as the dataset in Figure \ref{fig:lared}. In these datasets, audio recordings are commonly not available, due to the technical and logistic difficulty, and privacy challenges when equipping each study participant with a microphone. In-lab studies of the body movements of laughter have also often opted for video-only annotation of laughter, to align with the target task under study.



Despite the application of video-only labelling, it is unknown if video labelling of laughter has a relevant effect on the quality of annotations, and how annotations acquired in this way differ from the more common audio-based and audiovisual annotations. The same is true for audio-based labeling. Although works on audiovisual laughter detection generally perform annotations audio-visually, the benefits of including video at annotation time have not been verified. In other words, the consequences of the choice of annotation modality have received little attention in research. It is unknown, for example, how modality affects the ability to detect true episodes of laughter (ie. recall), and only true episodes (ie. precision), and how it affects measures of inter-annotator agreement, traditionally used to assess annotation quality. Furthermore, it is unclear whether ratings of intensity of laughter are congruent across modalities, a question with direct implications in the interpretation of laughter \cite{Mazzocconi2020}. Finally, the effect of labeling modality on model performance remains entirely unexplored. Answering these questions is important both for the interpretation of previous work focusing on a single modality, and for informing annotation choices in future work. We expect that this line of inquiry could lead to a better understanding of the multimodal perception of laughter, leading to a more holistic operationalization of laughter.     

In this work, we take a first step in this direction by studying differences in perception of laughter across modalities, with a focus on annotation task. First, we investigate how human ability to detect, segment and estimate intensity of laughter (three foundational tasks in laughter work) differ with and without access to video and audio. We made use of a free-standing interaction dataset collected during a real-life event, and containing naturally-occurring laughs (Figure \ref{fig:lared}). Laughter manifestations on body movement (eg. shaking, swaying, arm and feet movements) can be observed in the videos, but access to facial features is limited, due to occlusion and self-occlusion. These factors, along with the diversity of camera angles, and distances to the camera make the in-the-wild setting one of the most challenging scenarios for laughter perception, especially from video. 

Second, we study how labels acquired under different modality conditions affect the performance of machine learning models for laughter detection, segmentation and intensity estimation. We pay special attention to the question of whether video-acquired annotations result in performance comparable to that of audio and audiovisual annotations. Naturally, the input modality of the model plays an important role here. To comprehensively answer our question, we compare single modality (audio, video) and multimodal (audiovisual) models, for the different labeling modalities. In addition to video, audio and audiovisual inputs, we made use of accelerometer readings from chest-worn wearable devices worn by many participants in our dataset. Such accelerometer readings have been used in previous work for the detection of multiple social actions such as speaking \cite{Gedik2017, Vargas2019a, Raman2022b}, with performance competitive and often superior to that of video. This is likely due to their ability to capture subtle core body movements. Furthermore, wearable accelerometers have privacy and scalability advantages due to their low cost and their ability to capture information from the device wearer alone. Due to these advantages, we included acceleration readings as an additional model input in our machine learning analysis. We hypothesized that acceleration would have a behavior similar to video, since both modalities capture primarily body movement information. However, we expected acceleration to better capture laughter intensity when compared to video due to its orientation invariance, and to it not being affected by occlusion and cross-contamination like video is.

Our contributions are the following:

\begin{itemize}

\item We present the first human study of laughter annotations across annotation modalities, comparing between three conditions of interest in previous work: audio-only, video-only and audiovisual. We studied the three annotation tasks of laughter detection, time-localization and intensity rating.

\item We present a cross-modal analysis of annotations via inter-annotator agreement within and between conditions. We confirmed that agreement scores are significantly lower in the video condition, compared to when audio is present. Higher within-condition agreement scores than between-condition scores suggest that a different \textit{concept} was perceived across conditions.

\item We investigated the effect of the annotation modality in machine learning model performance. Mirroring the human study, we analyzed detection, intensity estimation and time-localization models. We trained and evaluated state-of-the-art models for different combinations of input modalities (audio, video, acceleration), training label modality (video, audio and audiovisual), and testing label modality (video, audio and audiovisual). Our results showed that, while audio and audiovisual models are sensitive to training label modality, models with video and acceleration inputs showed indistinguishable performance regardless of training label modality. This suggest that, despite the lower inter-annotator agreement of video-based labeling, it may be entirely appropriate to train models for laughter detection from body movements using video-acquired labels.

\end{itemize}


\section{Background and Related Work}
\label{sec:related}

In this section, we start by briefly introducing the research landscape surrounding laughter (Section \ref{sec:bg_laughter}). In section \ref{sec:bg_laughter_annot} we introduce work on laughter annotation and perception, including some work similar to ours. This section also provides relevant background to understand the design decisions behind our human annotation study. In section \ref{sec:bg_comp} we provide a similar overview of automatic laughter detection and related machine learning task, relevant for the decisions behind our machine learning experiments. 
 
\subsection{The Study of Laughter in Interaction}\label{sec:bg_laughter}

Laughter has been approached from the perspective of multiple scientific disciplines. Psychology, is concerned with, among others, the semantics and functions of laughter in interaction \cite{Ginzburg2015, Oatley2014, Scherer2009a}. In biology, the evolutionary origins \cite{Gervais2005} and physiological effects of laughter \cite{Miller2009} are subject of study. Meanwhile, social signal processing, speech and human-agent interaction fields are concerned with automatic tasks such as laughter detection \cite{Gillick2021, Petridis2013a}, classification \cite{Griffin2013, Griffin2015} and synthesis \cite{Silvervarg2012} , that enable artificial agents and models to consider human laughter and/or produce it \cite{Petridis2015}, with datasets being created for the study of laughter in specific \cite{McKeown2015b, Petridis2013, Jansen2020}.

Laughter is most often analyzed as a meaningful signal in the context of social interaction, as laughter is an overwhelmingly social phenomenon, found to be about 30 times more likely in social situations than when by oneself \cite{Provine2000}. To this end, drawing a parallel with the study of speech, Mazzocconi et al. \cite{Mazzocconi2020} distinguished two broad levels for the study of laughter: 1) laughter form and context and 2) laughter's (social) meaning and function.

Laughter form includes the physiology and body movements of laughter, and its acoustic features; and laughter context includes its positioning with respect to speech, to others' laughter, and to its object (the laughable). Most of the work on the form of laughter is concerned with it's phonetics and acoustic structure, with different coding schemes for segmentation of laughter into its constituent (acoustic) components often being used \cite{Truong2019}. Body movements occurring below the face are largely disregarded in the study of laughter form. Laughter intensity has also received attention as a dimension of laughter form \cite{Mancini2012, Niewiadomski2012, McKeown2013, DiLascio2019, Niewiadomski2015, Haddad, ElHaddad2018, Curran2018, Vernon2016}. Most laughter in conversations has been observed to occur at relatively low intensity \cite{Vernon2016}.

Laughter form and context influence a second level of analysis: the meaning and function of laughter, concerned with the semantics and the effects of laughter in an interaction. Here, Mazzocconi et al. proposes the following as the meaning of laughter: "The laughable \textit{l} having property \textit{P} triggers a positive shift of arousal of value \textit{d} within \textit{A}’s emotional state" \cite[p.4]{Mazzocconi2020}, where \textit{A} is the producer of the laugh. Mazzocconi et. al. \cite{Mazzocconi2020} link the value of the shift in arousal \textbf{d} directly to laughter intensity, highlighting its importance to laughter meaning. In fact, higher-intensity laughter has been observed to be strongly associated to humor use or humorous situations \cite{Vernon2016}. Despite the importance of laughter intensity in determining laughter meaning and function, it is not known to what extent perception of laughter intensity is congruent (or not) across modalities.

Laughter has been found to serve a multitude of functions. At the turn taking and coordination level, laughter can take the role of punctuation \cite{Provine1993, Provine2000} and can be used as a cue for topic termination / continuation \cite{Holt2010, Odonnell-Trujillo1983}. At the social level, laughter is commonly associated with expressions of enjoyment, affiliation, aggression, social anxiety, fear, comicality and ludicrousness, among others \cite{Poyatos1993}.  At the biological level, longer-lasting physiological changes with possible effects on our health  have been found as result of laughter\cite{Miller2009}.



\subsection{Automatic Laughter Detection, Classification, and Intensity Estimation}\label{sec:bg_comp}

Most research in laughter detection and classification focuses on the audio modality, where spectral features, pitch and energy features and voicing features are useful in discriminating laughter from speech \cite{Truong2005, Truong2007}. Multiple attempts have been made in the field of audiovisual laughter detection. The video modality generally corresponds to frontal videos of the face of the participant, possibly including the upper body. In a series of papers, Petridis et al. investigated audiovisual  laughter detection and discrimination \cite{Petridis2008,Petridis2008,Petridis2011}. Authors used a combination of static and dynamic features from both modalities to classify laughter episodes extracted from the AMI meeting corpus \cite{Petridis2008}.

To the best of our knowledge there have been no attempts to automatically detect laughter exclusively from the video modality. However, Mancini et al. \cite{Mancini2012} proposed an algorithm to automatically measure laughter intensity from the movement of shoulder and head keypoints in a video. Detecting actions from body movement has however been explored in the case of speech \cite{Beyan2020, Gedik2017}.

Previous work using full body poses as input, \cite{Griffin2015, Niewiadomski2016} showed that traditional classifiers are capable of recognizing and classifying elicited laughter from pose time series alone. Cu et al. \cite{Cu2017} classified laughter into two affective categories from body keypoints, using different classifiers.

\subsection{Laughter Perception and Annotation}\label{sec:bg_laughter_annot}

At its lowest level, laughter annotation is concerned with the recognition and segmentation of the form of laughter. Most of the work on the form of laughter is concerned with it's phonetics and acoustic structure. Laughter is typically classified in voiced and unvoiced laughter, depending on the degree of engagement of the vocal chords \cite{Petridis2015}. Speech-laughter is often used as a third formal category to indicate when speech and laughter intermingle \cite{Haddad2015}. Regarding its temporal extent, there is not a widely-accepted definition of what constitutes a laughter episode. Most studies of laughter delving into its structure have relied on audio waveforms for the segmentation of laughter, typically into laughter syllables or vowels (ha) at the lowest level, followed by bouts (sequences of syllables), which are separated by inhalations \cite{Trouvain2003}. Truong et al. propose a multi-level segmentation scheme to describe the structure of laughter \cite{Truong2019}. This scheme, however, relies on audio alone.

Body movements, especially those occurring below the face, have been largely disregarded in the study of laughter form. There are however, notable exceptions. In a perception study, Griffin \cite{Griffin2015} showed that humans are capable of recognizing laughter and even of classifying it functionally based on stick figures. The use of stick figures provided a way to isolate the body movement component of laughter. Note however, that in contrast to our work, this study was not concerned with annotation (where the goal is to use the most reliable information available) and did not analyze agreement with other modalities like audio.

In the work most similar to ours, but focused exclusively on facial movements, authors created visual, audio, and audiovisual laughter stimuli/examples from face recordings \cite{Jordan2010}. The audio contained different levels of artificial noise to make laughter more difficult to detect. 20 annotators indicated if they perceived laughter or not in these examples. The goal was to study how much the face contributes to the perception of laughter. The study reported that "visual laughter consistently made auditory laughter more audible" (ie. audiovisual laughter was easier to detect than audio-only laughter), a phenomenon also observed previously for speech perception \cite{Jordan2010}. Although this is, to the best of our knowledge, the only work to perform a cross-modal analysis of laughter perception, its findings do not necessarily generalize to our setting, where the video modality contains overall body movement information, but facial movements are not consistently available. Furthermore, being a perception study, they considered expert annotations to be ground truth, but provided no analysis of inter-rater agreement.


Most studies of automatic laughter detection (see Section \ref{sec:bg_comp}) rely on laughter annotations made from audio \cite{Petridis2008, Petridis2013a} (possibly automatic like the ones in AMI \cite{Carletta2007} and SEMAINE \cite{McKeown2012}) or audiovisually \cite{Reuderink2008, Mancini2013, Niewiadomski2016}. However, studies concerned with the body movements of laughter often obtained ground truth annotations from the video modality alone. \cite{Mancini2012} rated laughter intensity from body movements alone. Cu et al. \cite{Cu2017} annotated five affective categories of laughter from body movement, without sound. These studies, however, do not offer a comparison with audio-based annotation, and it is therefore uncertain to what extent annotations would be congruent across modalities.

\section{Our Approach}
\label{sec:approach}

Answering our research questions requires the annotation of a large set of laughter segments with associated audio and video signals. Measuring inter-annotator agreement across conditions additionally requires that the same segments are annotated by multiple annotators. Annotations must also be done by a representative sample of annotators, large enough to ensure that individual biases do not drive the results. Ideally, a large number of annotators would each watch our complete dataset (with more than $50h$ of individual behavior) to find and annotate laughter episodes. This, however, would involve hundreds to thousands of hours of human labour. Due to the relative scarcity of laughter in conversation, most of this time would be spent listening to speech with only sporadic laughter. Therefore, the first simplification that we adopted was to pre-localize \textit{laughter candidates}. \textit{Laughter candidates} are segments where the author of the study (who did the pre-annotation) perceived laughter to occur. The pre-annotation was done inclusively, meaning that in case of doubt laughter was always annotated. These positive candidates were complemented with negative examples, where laughter was not perceived to occur, to obtain a dataset of laughter / non-laughter candidates. Positive candidates were expanded temporally to include more context than just the laughter episode, resulting in segments of $~7s$ in length.

\begin{figure*}
\centering
  \includegraphics[width=0.8\textwidth]{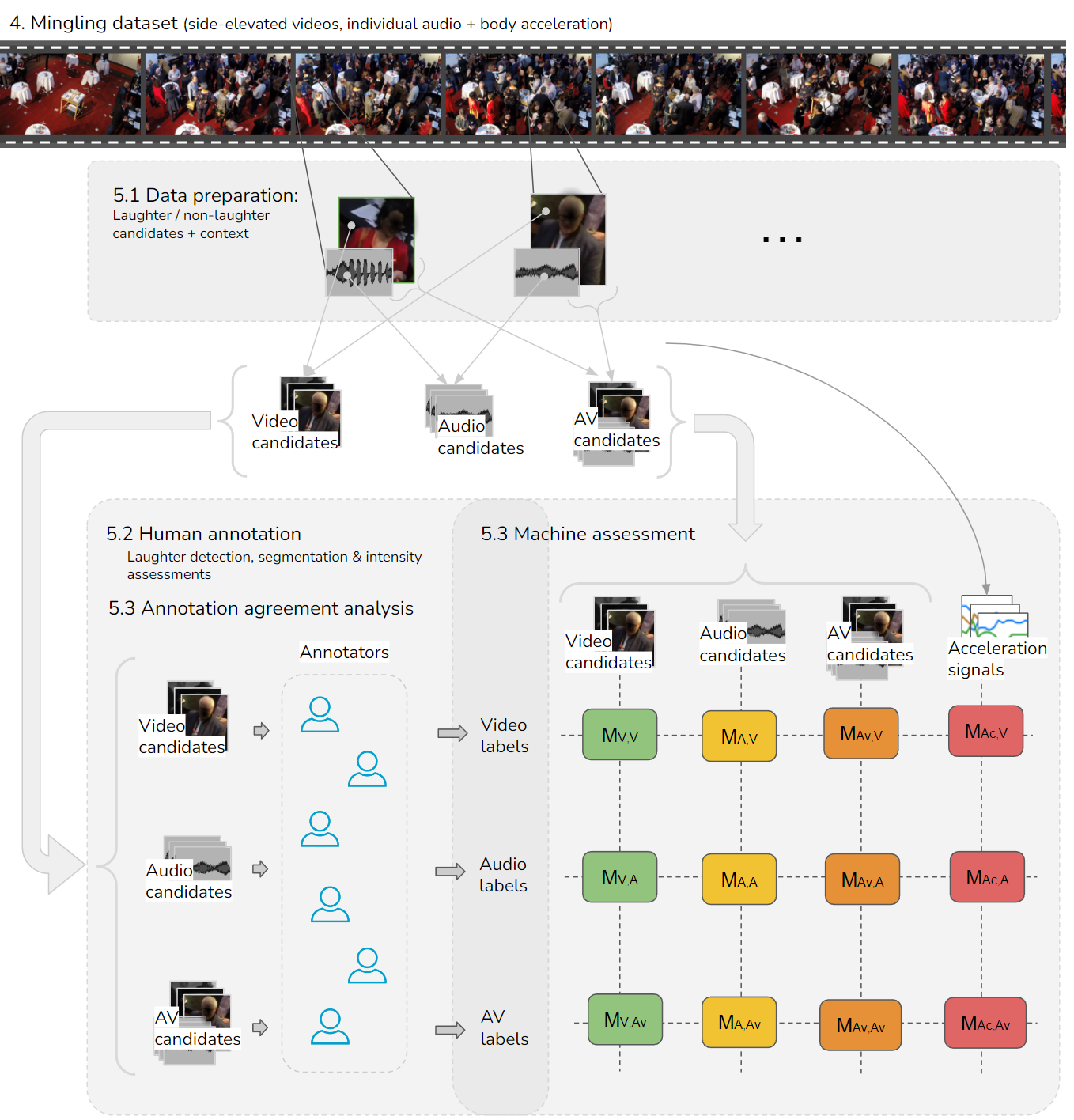}
  \caption{Overview of our study. From a mingling dataset with video, individual audio, and accelerometer readings (Section \ref{sec:dataset}), we extract pre-annotated segments of potential laughter and speech, each of $~7s$ in length. These segments are annotated for laughter presence, segmentation and intensity under three conditions: audio-only, video-only and audiovisual. We analyze the labels directly (Section \ref{sec:results_agreement}) and use the different sets of labels to train and benchmark models for laughter detection, segmentation and intensity estimation (Section \ref{sec:results_comp}).}\label{fig:study}
\end{figure*}

Figure \ref{fig:study} shows an overview of our complete study. In the following sections we detail our dataset and methods. In Section \ref{sec:dataset} we present the audiovisual dataset chosen, containing audiovisual recordings of laughter in a real-life mingling setting. In Section \ref{sec:methods} we delve into our methodology, starting with the design of our design of our human annotation study. Each candidate segment was annotated by multiple annotators, across three annotation conditions: audio-only, video-only and audiovisual (Section \ref{sec:method_annot}). We performed a direct analysis of annotator agreement in the obtained annotations (Section \ref{sec:method_agreement}), focused on measuring differences across modalities.

Finally, the same labels were used in a computational analysis of the impact of labelling modality in machine learning model performance (Section \ref{sec:method_comp}). We analyzed the automatic tasks of classification, segmentation and laughter intensity regression.




\section{Dataset}\label{sec:dataset}

Our dataset was collected during a networking event in Delft, The Netherlands. Participants in the experiment were members of a business networking group, organizing regular events. During the event, most of the interaction consisted of free-standing conversation (Figure \ref{fig:lared}). Participants were free to move around and talk as they pleased. Several such \textit{mingling} datasets have been published \cite{Alameda-Pineda2015, Cabrera-Quiros2018a}, but our dataset was the first to contain high-quality individual audio recordings, opening the door for cross-modality studies such as this one. Because many participants were acquainted with each other and this was a special event commemorating an anniversary of their organization, conversations were mostly friendly, and laughter abounded. Drinks and snacks were served. The event also included several pre-planned activities, where participants were expected to play a social game or watch  performance. We excluded these moments and made use only of the segment of the data containing free, unscripted interaction. The following data was collected during the interaction:

\begin{LaTeXdescription}
	\item[Body Acceleration.] A custom-made wearable accelerometer sensor that was hung around the neck and rested on the chest like a smart ID badge measured upper torso acceleration.
	\item[Individual Audio] Lavalier-type microphones attached to the face of participants via Lavalier tape recorded sound at 44.1kHz. Microphones were connected to a Sennheiser SK2000 transmitter worn around the waist area. Our audio equipment consisted of 32 such microphones. These individual audio recordings were used to obtain Voice Activity Detection (VAD) labels at $100~Hz$ for each participant, making use of rVAD \cite{Tan2020}, a state-of-the-art unsupervised VAD method specially designed for noisy audio. $100~Hz$ is the fixed output frequency of rVAD and enough to capture even single syllables in languages like English \cite{Crystal1990}.
	\item[Video] 12 overhead cameras and four side-elevated cameras were placed above and in the corners of a video zone. Participants were informed about this video zone, and asked to stay outside if they did not wish to be recorded. In this work we only make use of the side elevated cameras, due to it being a viewpoint more familiar to observers and able to capture the face. Figure \ref{fig:lared} shows a capture of the four elevated camera views.
\end{LaTeXdescription}

In coordination with event organizers (because this was a celebratory event not primarily planned for data collection), it was decided that each participant would be free to choose which sensors to wear (microphone, accelerometer, or both). Of about 100 attendees to the event, 43 consented to wearing a microphones or accelerometer sensor. Of them, 20 were male and 23 female. The rest of the subjects decided not to take part of the data collection, or could not be given a sensor due to our supply limit. Most interactions took place within the video zone.

\section{Methods}
\label{sec:methods}

In this section, we detail the methods used in our study of laughter annotation: 1) our method for obtaining laughter / non-laughter candidates for annotation, 2) our choice for annotation method, 3) the study design (ie. assignment of laughter candidates to annotators, and related decisions), and 3) a description of the annotation process followed by each annotator.

\subsection{Laughter candidate generation}\label{sec:candidate_gen}

To obtain a set of laughter candidate segments (thin slices) to be annotated in our human study, the first author localized any \textit{possible occurrences} of laughter in the dataset by watching the audiovisual recordings for every data subject and segmenting perceived laughter episodes using the annotation software ELAN \cite{Elan2021} by indicating the start and end of each laugh on top of the audio waveform. We were deliberately inclusive in incorporating segments when in doubt. Annotations closer than $1s$ apart were considered a single laughter episode. Not all segments, however, were visible in the videos. Therefore, we additionally annotated the cameras, if any, in which a particular laughter episode was visible. Segments present in multiple cameras were only considered once by randomly picking one of the camera views as its \textit{audio modality}. Segments not present in any camera were discarded such that only segments with both audio and video were used in the experiments.

\subsubsection{Negative candidate generation}
In addition to these, we automatically extracted a number of negative segments likely containing no laughter from the rest of the dataset. Due to the fact that our conversing groups were often large, a given subject would be a listener in their conversation for most of the time. To avoid having mostly segments of \textit{listening behavior} in this negative set, we sampled negative candidates from speech utterances as given by our VAD labels. Additionally, since we observed important differences in laughter frequency between subjects, we sample the number of negative samples per subject proportional to the distribution of positive samples. In this way, the subject distribution is the same for laughter and non-laughter candidates. Concretely, our sampling procedure for negative samples:

\begin{enumerate}
    \item samples a subject $S$ with probability $P_L(S)$ where S is the probability of a positive laughter candidate belonging to subject $S$. 
    \item samples a speech utterance uniformly from the set of speech utterances of $S$ of length $l_{min} < l < l_{max}$ where $l_{min}$ and $l_{max}$ are the lengths of the shortest and longest laughter episodes. Although we don't precisely follow the distribution of laughter lengths, we found this solution to be sufficient given that the goal was simply to avoid very long speech utterances from being introduced as negative examples. We do not expect the length distribution to have a biasing effect on human subjects.
\end{enumerate}


\subsubsection{Expanding candidates in time}
\label{sec:ctx_time}

Finally, laughter candidates (positive and negative) were expanded in time. The goal of expanding in time was to more closely resemble the process of annotating laughter in the wild, where it is unknown when laughter might happen, and at the same time allow some time for the annotator to process temporal context in the scene. To this end, we expanded each segment at both ends with a duration randomly (uniformly) sampled between $1.5s$ and $3.5s$. we set the bottom of this range ($1.5s$) to be close to the mean length of a laughter episode. Empirically, this was enough to process the scene and be ready for annotation. We set the top of the range based on the goal of obtaining candidate lengths below $10s$ to maintain the annotation process fast. We used a uniform distribution (maximum entropy) to minimize predictability of the location of the (potential) laughter episode.

\subsubsection{Spatial localization via bounding box}
\label{sec:ctx_space}

\begin{figure*}[ht!]
\centering
\begin{subfigure}{.49\textwidth}
  \centering
  \includegraphics[width=\textwidth]{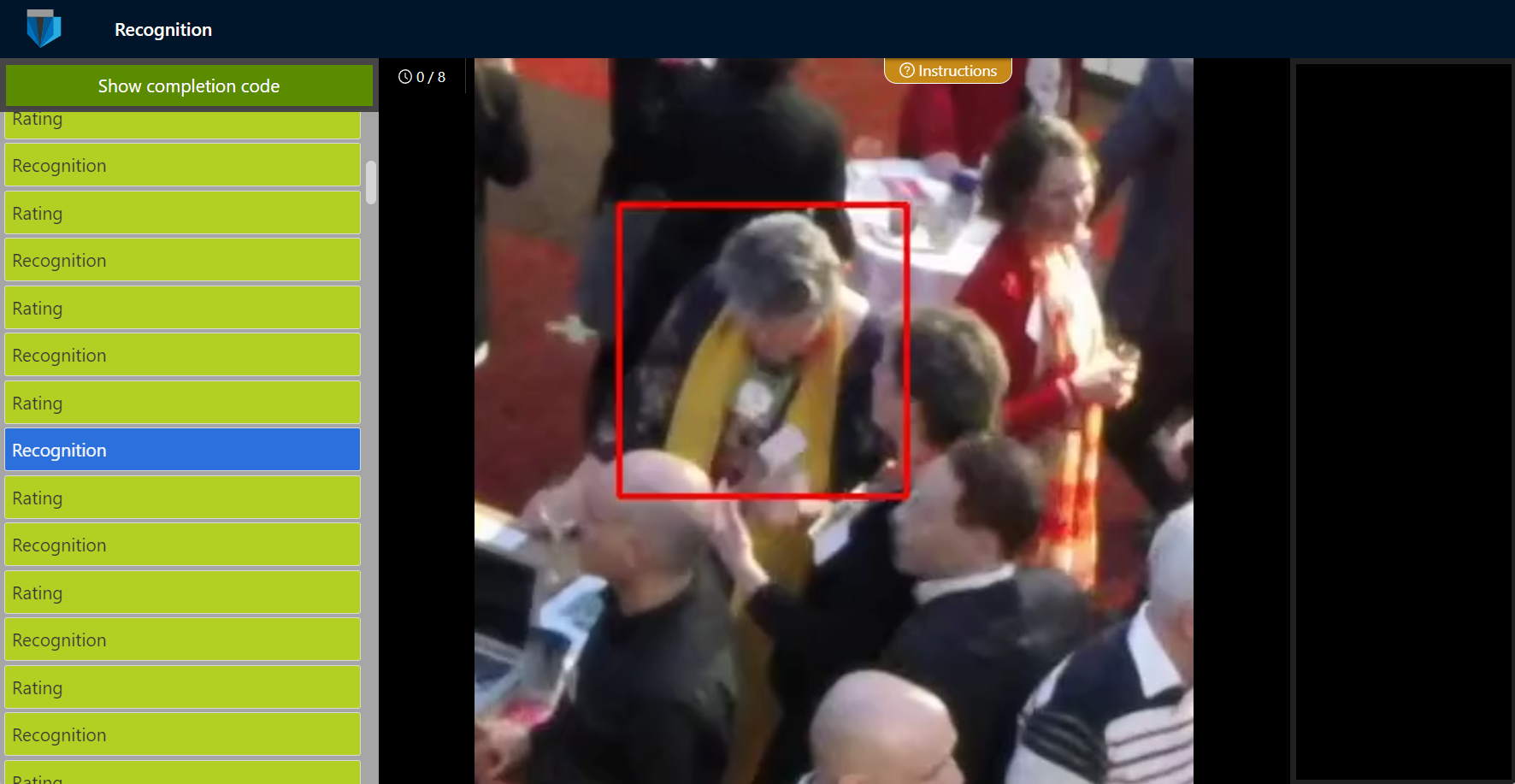}
  \caption{Recognition and continuous annotation step.}
  \label{fig:covfee1}
\end{subfigure}\hspace{.01\textwidth}%
\begin{subfigure}{.49\textwidth}
  \centering
  \includegraphics[width=\textwidth]{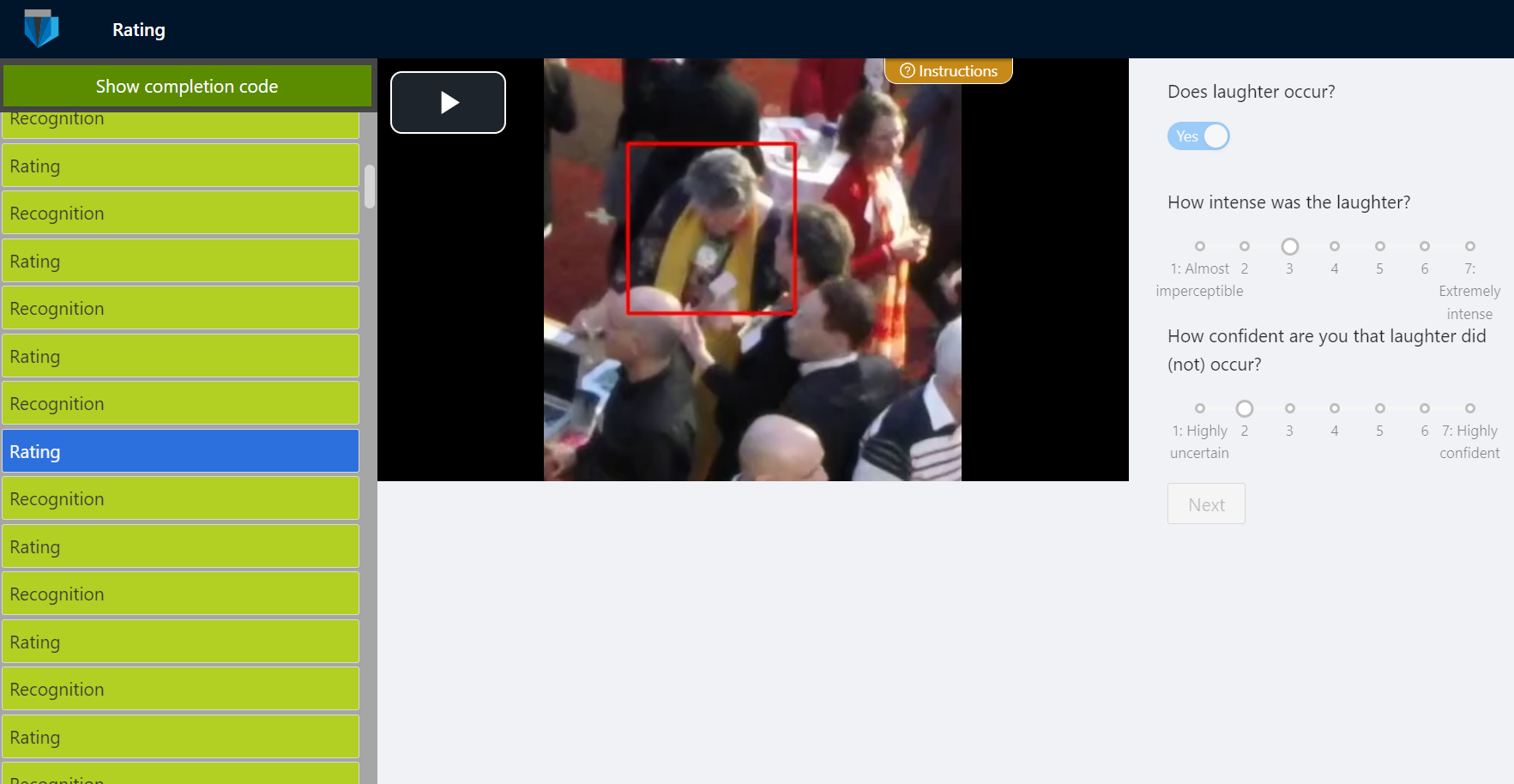}
  \caption{Intensity and confidence annotation.}
  \label{fig:covfee2}
\end{subfigure}
\caption{Screenshots of the annotation interface in Covfee \cite{Quiros2022}. The two steps shown were repeated for every example that an annotator rated. In (a) annotators were shown a target person marked by a red box, and part of the scene around the target, and instructed to hold down a key when laughter was perceived to be occurring by the target person. The interface provided visual feedback when the key was held down. In (b), subsequently, annotators rated laughter intensity (Likert scale 1-7) and their confidence in their assessment (Likert scale 1-7). They were able to replay the segment if they desired. Detailed instructions were provided to annotators on how to perform both steps. For more details, see Appendix \ref{app:annotation}}
\label{fig:covfee_iface}
\end{figure*}

Since our side-elevated camera views captured most of the interaction scene, but laughter annotations needed to be for a single target subject at a time, this person needed to be indicated to annotators. This was done by annotating a single bounding box around the target person for the first frame of the video. The box was drawn tightly, to contain all visible parts of the target person, but not more visual context than that. This box would be shown to the annotators for the frame of the video in the annotation study (see Figure \ref{fig:covfee1}). However, since participants in a conversation often laugh together, the laughing status of neighboring participants is a useful indication of the target person's status. To allow annotators to use such kind of contextual information, the video provided to them was cropped beyond the borders of this bounding box. To account for differences in distance to camera (and this pixel size of the participants in the video) we expanded the bounding box relative to its own size by multiplying its width and height by 3 (constrained to fitting within the frame) and maintaining its center. Our observations showed that this was in most cases enough to capture the interlocutors of the target person, while maintaining good visibility of them.

\subsection{Annotation of laughter candidates}\label{sec:method_annot}

Central to our study of laughter annotation is a correct design of the process to be followed by annotators to rate laughter. We decided to evaluate the task of laughter localization (in time), since this a canonical first task in the study of laughter in the wild. After localization, laughter may be further segmented, classified in multiple ways, or rated for intensity.

The process of localization involves both perception of the target media, and some form of input into a computer interface in response to it. Regarding perception, decisions include: what modalities to make available to the annotator, and the types and amounts of context available to them (see Sections \ref{sec:ctx_time}, \ref{sec:ctx_space}).

Regarding input techniques, actions in videos are traditionally localized using tools such as ELAN \cite{Elan2021}, where the user localizes the start and end frame of the action, which is then annotated by drawing an interval on top of the audio waveform. In tools such as Vatic and CVAT, actions are annotated via flags, which are turned on for the frame when the action is deemed to start, and off for the end of the action. In affective computing, \textit{continuous annotation} techniques are commonly used to annotate continuous-valued variables such as arousal and valence. In \textit{continuous annotation}, annotators control the value of the target variable while watching the subject in video, usually without pause. This has the advantage of letting the annotator perceive the behavior without interruption, and being efficient and predictable in terms of time needed to annotate. On the other hand, continuous techniques require the annotator to maintain attention in the video, which can be mentally exhausting for long periods of time. Continuous methods also necessarily introduce a reaction time delay. Multiple techniques have been proposed to mitigate these delays.

We chose to make use of continuous annotation for our study due to the mentioned advantages. Since our laughter candidates are only $~7s$ long and annotators can pause between candidates, we do not consider maintaining attention in each candidate to be a big challenge. We mitigated reaction delay by making use of an experimentally defined offset, as detailed in Section \ref{sec:delay_corr}. We made use of a binary action localization technique implemented in the Covfee framework \cite{Quiros2022}, that asks annotators to hold down a keyboard key when they perceive laughter to be occurring. Its graphical interface is shown in Figure \ref{fig:covfee1}. This process allows annotators to maintain focus on the phenomenon and context by minimizing the input effort, while still giving us access to high-resolution segmentations of laughter. Since the annotation time is shortened and predictable, this process also allowed us to obtain more annotations per annotator, relevant for our study design (Section \ref{sec:study_design}).

After the continuous annotation step, for each candidate, we asked annotators explicitly whether they perceived laughter to occur, their perceived laughter intensity, and their confidence in their laughter ratings (Figure \ref{fig:covfee2}). Annotators could replay the laughter episode if they desired.

\subsubsection{Crowd-sourced Annotation Process}\label{sec:study_design}

As introduced under Section \ref{sec:methods}, answering our research questions requires annotations of laughter under three conditions: audio-only, video-only and audiovisual. Measuring agreement within a condition imposes the requirement that at least two annotators rate each $(candidate, condition)$ combination, ie. at least 6 annotators rate each candidate. A sufficient number of candidates must also be annotated to be able to train our computational models and measure agreement over a large-enough set. Finally, for access to a large pool of annotators, annotations would be crowd-sourced and each annotation HIT (human intelligence task) should ideally not last longer than approximately 45 minutes to avoid mental fatigue. This imposes an upper bound on the number of samples that a single annotator can rate. We estimated this upper bound to be around 90, taking 30 seconds of annotation time per candidate, given the tasks that we asked from annotators for each candidate (Section \ref{sec:method_annot}). One other natural choice to consider was whether to use a between-subjects (each annotator rates in one condition) or within subjects design (each annotator takes part in all three conditions).

To maximize the number of annotators per condition, we opted for a study where each annotator takes part in all three conditions. To avoid bias, we impose the restriction that one annotator never annotates the same candidate under different (or the same) conditions, ie. one annotator rates three disjoint sets of candidates. Additionally, we set a maximum total annotation time per annotator to be around 45 minutes. Through preliminary tests we determined that this time was comfortably enough for each annotator to annotate 84 candidates, 28 in each condition. 

\begin{figure}
\centering
  \includegraphics[width=1\columnwidth]{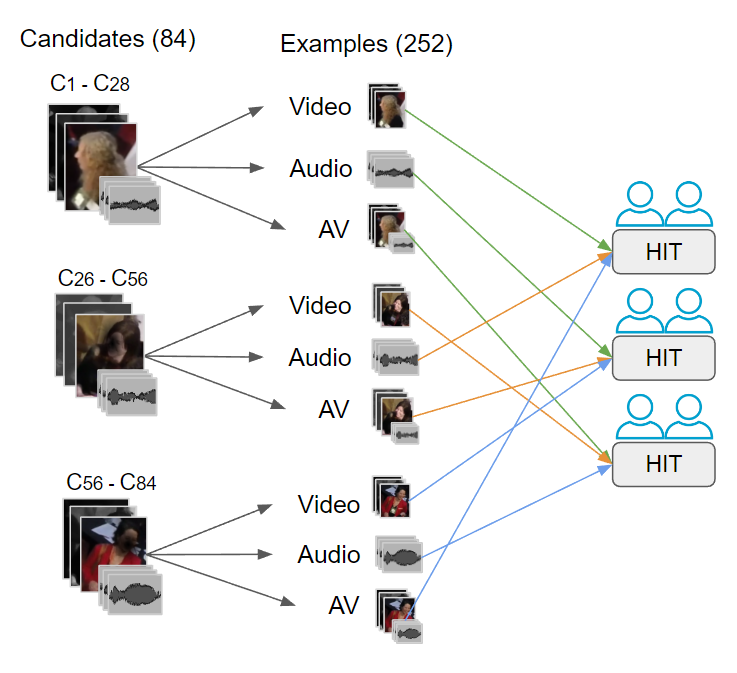}
  \caption{Structure of the annotation stage of our study. Sets of 84 randomly-selected candidates are separated into 3 equal-size sets of 28 candidates. Candidates are then separated into their audio and audiovisual modalities and assigned to HITs such that each HIT contains 28 distinct candidates per condition. Each HIT was annotated by 2 annotators.}\label{fig:hits}
\end{figure}

According to these design decisions, we divided our 659 candidates into 7 sets of 84 examples and one set containing the remaining 71 candidates. Each of these candidates sets was in turn divided into three equal-size subsets (for the three conditions). Each permutation of these three subsets resulted in a different human intelligence task (HIT), each containing the same candidate subsets but mapped to different conditions. Figure \ref{fig:hits} is a diagram of this process for each set of 84 candidates. Each HIT was completed by two annotators. Annotating all candidates required 48 annotators in total. This design allowed us to compute pair-wise inter-annotator agreement (per condition) over sets of 28 paired ratings, for 24 distinct pairs of annotators.

The following section explains the details of the annotation process, including the structure  of each HIT.

\subsubsection{Annotation HITs}\label{sec:the_hit}

To ensure access to a large pool of annotators, we crowd-sourced our annotations online via the Prolific crowd-sourcing platform \cite{prolific}. We implemented the annotation flow using the Covfee annotation tool \cite{Quiros2022}, which supports both continuous annotations and non-continuous ratings. Each annotation HIT consisted in a series of tasks to be followed in order. In general, each HIT contained several introductory tasks and examples, followed by three annotation blocks, one for each modality condition. The order of video-only, audio-only and audiovisual blocks was randomized to avoid ordering bias due to factors like fatigue. The ordering of laughter examples within each block was also randomized for the same reason. The detailed structure of each HIT is presented in Appendix \ref{app:hit_struct}.

\subsubsection{Annotation Delay Correction}
\label{sec:delay_corr}

Delays in continuous annotation have been investigated within the affective computing community for continuous-value annotations of affective dimensions. Mariooryad et al. proposed a model for correcting delay in annotations by maximizing mutual information between the continuous label time series and an auxiliary signal containing facial keypoints \cite{Mariooryad2015}. However the authors also showed that simply offsetting annotations by a constant value most often resulted in performance as good as that of their method. Other works have proposed models that are robust to annotation delays \cite{Huang2015a, Khorram2019}.

Despite these works, it is unclear to what extent delay may be task-specific (ie. how strongly it depends on the particular actions being annotated). We therefore decided to measure delay directly for our task and annotators. At six points in each annotation HIT (two per condition, see Section \ref{sec:the_hit}), we inserted special \textit{calibration} (positive) laughter examples, which were the same for all annotators. We precisely labeled the onset and offset times of laughter in these six examples, using ELAN \cite{Elan2021}. This allowed us to calculate a delay in the annotator's continuous labels, to approximate the average delay of each annotator. We used this average annotator delay as correction offset for an annotator's labels.

\subsection{Measuring Inter-Annotator Agreement}\label{sec:method_agreement}

Our annotation analysis involves both a) measuring differences between condition-specific annotations and expert-obtained pre-annotations of laughter and b) measuring differences across annotators, for paired samples. Our study design allows for the computation of inter-rater agreement, or reliability, within and across conditions. Cohen's Kappa, Fleiss' Kappa, and Krippendorfs Alpha are some commonly used measures of agreement. 

For nominal values (eg. laughter / non-laughter) Cohen's Kappa is capable of computing agreement between exactly two annotators. Although Cohen's Kappa is subject to biases in some instances, it still has been recommended by previous work for fully crossed designs with multiple coders, by computing the average of pairwise agreement \cite{Nilsson1960}. Some measures designed for non-fully crossed designs, like Fleiss Kappa, rely on the assumption that the raters assigned to each sample are randomly sampled from a large population of coders. Although our study design is not fully-crossed, each of our annotator groups rated a set of examples not rated by any other pair (ie. our study consists of a set of fully-crossed designs). We therefore decided to make use of the arithmetic mean of pairwise Cohen's Kappa values as a measure of agreement for nominal values.

Cohen's Kappa is however not appropriate for interval / ordinal values like laughter intensity or annotator confidence (Likert scale 1-7). Here, we used Krippendorff's alpha, a reliability measure applicable to any number of raters and which adjusts for small sample sizes. We similarly averaged pairwise Krippendorff's alpha values over rater pairs. Correlation coefficients are other commonly used tools in analyzing differences between interval or ordinal ratings. It should be noted, however, that correlation measures ignore differences in the expected magnitude of the ratings between annotators, and should not be considered measures of inter-rater reliability.

\subsubsection{Agreement in Action Localisation}

Measuring agreement between time series is a notoriously open challenge, as traditional measures like Cohen's Kappa are not designed with time series in mind. Straightforward application of these categorical agreement measures at the frame level is likely to result in higher agreement values the longer the actions are (annotator delay becomes negligible) but not provide an interpretable measure of agreement, as for categorical data. We decided to analyze agreement of our continuous binary annotations via computing intersection over union (IoU), computed as the size of the intersection of positive annotations over the size of the union of positive annotations. This is inspired on action localisation metrics used in computer vision \cite{Chao2018a}, which make use of IoU to identify true positives. An IoU of one indicates perfect correspondence between the positive sections of the time series, while an IoU of zero indicates no overlap. In cases where both signals contain only negatives (the union is undefined), we set the IoU to one to indicate agreement.

\subsection{Automatic, Laughter Detection, Intensity Estimation, and Segmentation}\label{sec:method_comp}

Video-based models for detecting, assessing (eg. intensity) and segmenting actions have been extensively study in computer vision and pattern recognition (Section \ref{sec:bg_comp}). We make use of modern approaches within these fields. Regarding the video modality, due to the small size of our dataset, training state-of-the-art methods would not result in good performance. We therefore focused on approaches with pre-trained models available to use as feature extractors. Among those, 3D convolutional neural networks (CNNs) are known to reliably achieve top performances in action recognition benchmarks. We decided to make use of a 3D ResNet pretrained on Kinetics-400, a large action recognition dataset with 400 action classes and over 300000 labeled video clips. The network implementation and models are available as part of the \textit{Pytorchvideo} library \cite{fan2021pytorchvideo}.

Regarding audio-based models, work by Gillick et al. \cite{Gillick2021} investigated laughter detection in two datasets with significant background noise. One of these, the Audioset dataset \cite{Gemmeke2017} is freely available to download. This dataset of 10-second clips from Youtube videos recorded in a variety of in-the-wild settings contains 5696 clips labeled as containing laughter. In their implementation, the authors provided a list of randomly-sampled clips to be used as negatives, for a total data set size. Given that this dataset had more examples and variety of subjects than ours, we decided to pre-train our audio-based model on it. We made use of the same model proposed by Gillick et al. \cite{Gillick2021}, a 2D ResNet model operating on the spectrogram of the audio inputs. We trained on 85\% of the dataset, with 15\% separated to determine a good stopping point. We otherwise used the same hyper-parameters used by the authors for their ResNet model without additional augmentations.

As motivated in section \ref{sec:approach}, we made use of acceleration as a modality capturing overall body movements. As acceleration-based model, we made use of ResNet variant for time series, implemented as part of the \textit{tsai} library \cite{tsai}. Given the much lower dimensionality of the acceleration data (compared to video and audio), and the lack of availability of similar whole-body acceleration datasets, we trained this model from scratch.

For both video and audio models, we used the pre-trained models as feature extractors by freezing all parameters and removing their network heads. For classification, the features output by the base networks (with dimensionalities: 2304 (audio), 8192 (video) and 128 (acceleration))  are fed into a head consisting in a linear layer followed by an output sigmoid layer and binary cross-entropy loss, standard choices for binary classification. For multimodal evaluation, we concatenate the features from multiple models before the head of the network.

We decided to approach intensity estimation as a regression task, given the interval nature of the ratings. We follow the same model structure, but we removed the sigmoid computation from the output and used L2, or mean squared error (MSE) loss, a standard choice for data with no outliers. 

\begin{figure}
\centering
\begin{subfigure}{.49\textwidth}
  \centering
  \includegraphics[width=\textwidth]{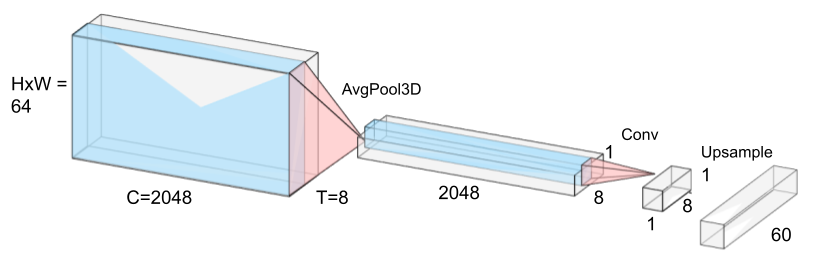}
  \caption{Time series ResNet head (acceleration modality).}
  \label{fig:seghead:accel}
\end{subfigure}

\begin{subfigure}{.49\textwidth}
  \centering
  \includegraphics[width=\textwidth]{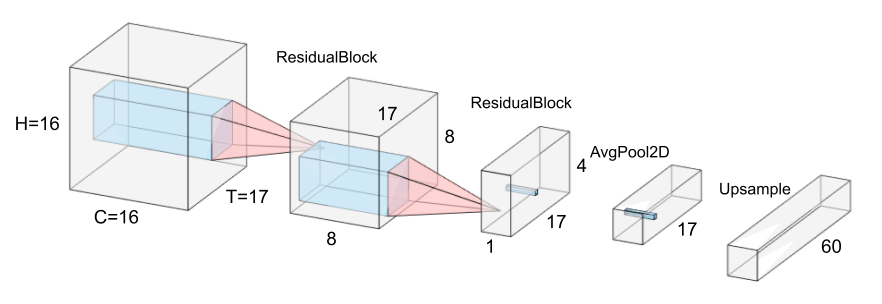}
  \caption{Audio ResNet head.}
  \label{fig:seghead:audio}
\end{subfigure}

\begin{subfigure}{.49\textwidth}
  \centering
  \includegraphics[width=\textwidth]{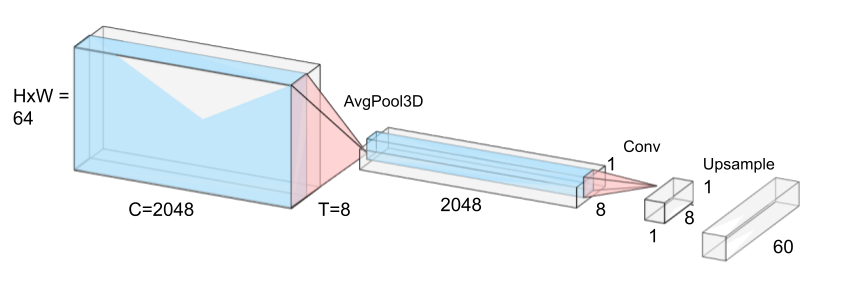}
  \caption{Video ResNet (slow model) head.}
  \label{fig:seghead:video}
\end{subfigure}
\caption{Segmentation heads for acceleration, audio and video models. The first block represents the feature map before the head of the ResNet model, for each modality method. Subsequent operations pool and convolve over the spatial and channel dimensions, and up-sample the time dimension to the length of the target segmentation mask (60).}
\label{fig:segheads}
\end{figure}

For segmentation, we decided to approach the task as the estimation of a binary mask (ie. of our continuous binary annotations). This would allow us to use the same base networks and pre-trained models. However, multimodal fusion should ideally not be done by concatenating features, since their time dimension encodes information likely useful for segmentation. We therefore implemented separate segmentation heads per modality, which are fused at the output via average pooling. Figure \ref{fig:segheads} shows the architecture of the implemented segmentation heads. For all models, we apply pooling and convolution operations over the spatial and channel dimensions, and up-sample the time dimension to the length of the target segmentation mask (60).

\subsubsection{Generating Train and Test Samples from Laughter Annotations}
\label{sec:model_windowing}

\begin{figure}
    \centering
    \includegraphics[width=0.7\columnwidth]{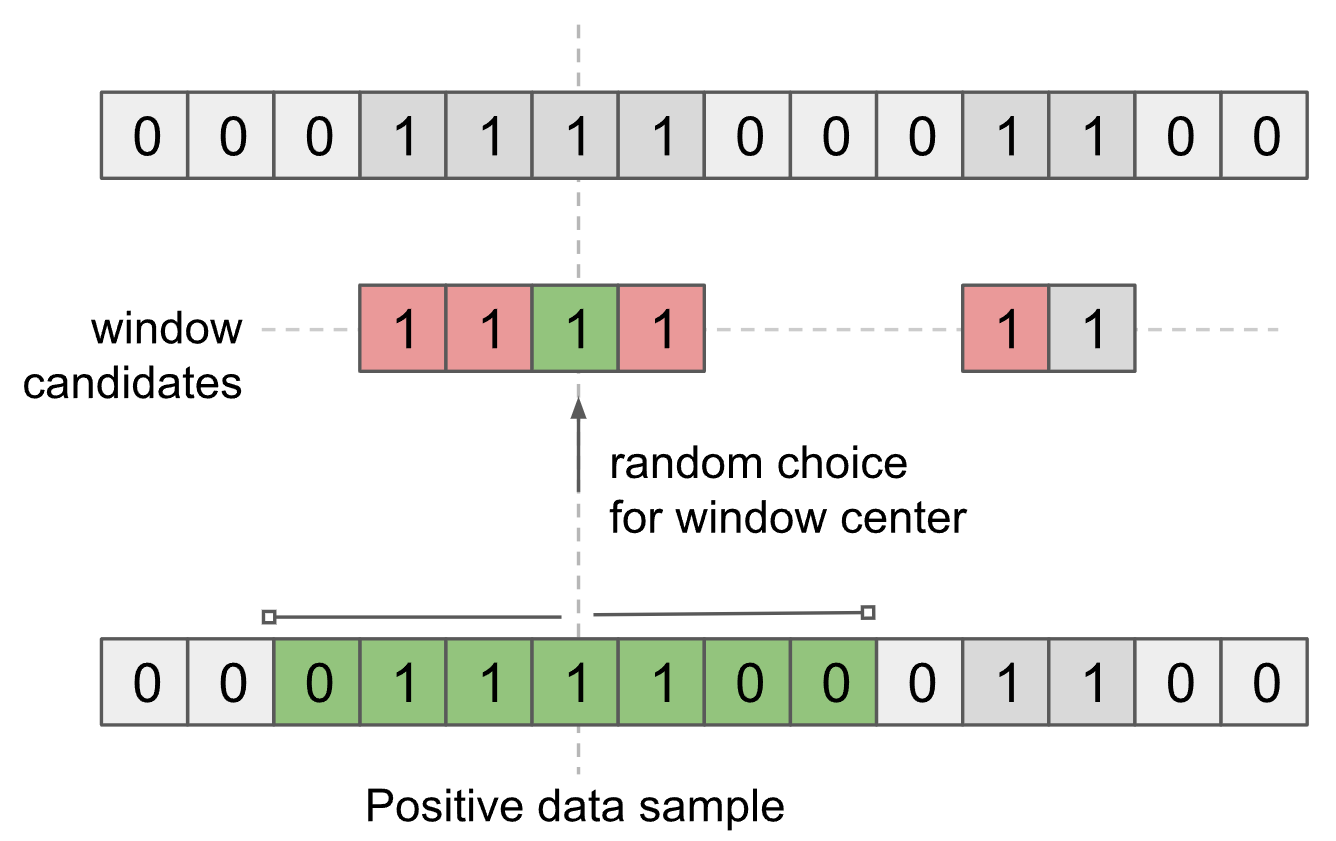}
    \caption{Illustration of the process used to select positive laughter samples for our machine learning tasks. Given the binary laughter / non-laughter annotations for a particular segment, we select a location for the window center from the positively-annotated intervals in the signal. We then extract a window of 1.5 seconds around the chosen center. We pad if necessary.}
    \label{fig:windowing}
\end{figure}

The first question we addressed regarding evaluation was how to obtain training/testing samples from our laughter annotations. Given that the examples seen by laughter annotators contained a significant amount of context, using the complete $~7s$ candidates would not be ideal, given the much shorter average duration of laughter. Furthermore, our models made use of fixed size inputs, and the examples rated by annotators were not fixed length. Continuous annotations were also free form, and could contain multiple detections of laughter within the same slice. To address the situation, we use the continuous binary labeling signal as reference, and sample short positive examples around its positive sections (ie. exactly where laughter was detected to have occurred). Figure \ref{fig:windowing} shows a simplified depiction of the process. Given a binary annotation signal with at least one positive segment, we consider the intervals within its positive segments as candidate window locations. We sample uniformly from these locations to select the window center, which determines the limits of the window. For negative examples (ie. with no positive segments), we consider every location in the signal to be a candidate for the window center (ie. we perform a random crop).

To determine the size of the window, we looked at the distribution of laughter lengths, as obtained from our continuous annotations. The average laughter length was 1.14 seconds, with a long-tailed distribution such that 80 percent of laughs were under 1.56 seconds. We chose a length of 1.5s. This length guarantees that most laughter segments will be contained in the window without excessive non-laughter context.

In evaluation, to avoid randomness, instead of the sampling procedure, the window is always centered on the laughter episode for positive examples. For negatives, the window is always in the middle of the complete segment.

The same sample generation process was followed for the three tasks  of laughter detection, intensity estimation and segmentation, although the labels differ per task. For detection, the sample is labeled positive when it comes from a positive annotation segment, and negative otherwise. For intensity estimation, the segment is labeled with the global intensity label (Likert scale 1-7) for the laughter candidate. To avoid reducing the size of the training set for the intensity estimation tasks, negative samples were included, and assigned an intensity of zero. For segmentation, the target is a vector corresponding to the continuous binary labels within the target window (at $30~fps$, resulting in a vector of size $45$ for our $1.5s$ windows).

Note that our annotation study involved two raters per candidate and condition. Both of these evaluations are included in the dataset, meaning that our models may be input the same examples twice in the same epoch. Note however that differences in the annotations and the randomness in our sample generation procedure mean that this is unlikely to be the case.

\subsubsection{Evaluation Procedure}
\label{sec:evaluation}

For evaluation, we make use of standard metrics for each task. For classification, we make use of the area under the ROC curve (AUC), a metric designed for binary classification and invariant to class imbalance. For regression, we make use of Mean Squared Error (MSE). We also make use of AUC for segmentation, where we treat every window element as one separate prediction. Although metrics like Intersection over Union (IoU) are more commonplace in segmentation, we made use of AUC due to it not being affected by class imbalance.

We evaluated via 10-fold cross-validation, to obtain an aggregated performance measure over the whole dataset. We  used default values for all model hyper-parameters. We used the first fold for tuning the number of epochs to train for (per combination of modalities) and excluded the first fold from evaluation.

\section{Results}
\label{sec:results}

\subsection{Comparison of Human Laughter Annotation Agreement Across Modalities}
\label{sec:results_agreement}


To test out hypotheses around differences in annotations across modalities, we started by calculating inter-annotator agreement within and across modalities. We did so via pairwise computation of agreement metrics (Section \ref{sec:method_agreement}). Our study design resulted in 24 valid (pairwise) comparisons within each condition, and 96 between each condition pair. Tables \ref{tab:detection_agreement} and \ref{tab:intensity_agreement} show the results of our agreement calculations for laughter detection and intensity rating. Standard deviations are shown in parentheses, and are calculated across pairs of annotators. Agreement scores for laughter detection (Table \ref{tab:detection_agreement}) show that the audio and audiovisual conditions have greater agreement between them, with video being significantly lower. The video condition had higher within-condition agreement than agreement with the other labeling modalities.

\begin{table}[hbt!]
\caption{Precision, recall and inter-annotator agreement measures across modalities.}
\label{tab:agreement}

\begin{subtable}{.5\textwidth}
\centering
\caption{Laughter detection inter-rater agreement (Cohen's Kappa)}
\label{tab:detection_agreement}
\begin{tabular}{llll}
\toprule
  Condition &    Audio-only &    Video-only &   Audiovisual \\
\midrule
 Audio-only & 0.823 (0.153) &  &  \\
 Video-only & 0.396 (0.186) & 0.550 (0.146) &  \\
Audiovisual & 0.795 (0.144) & 0.424 (0.183) & 0.805 (0.144) \\
\bottomrule
\end{tabular}
\end{subtable}

\vspace*{.3cm}

\begin{subtable}{.5\textwidth}
\centering
\caption{Laughter intensity inter-rater agreement (Krippendorff's alpha)}
\label{tab:intensity_agreement}
\begin{tabular}{llll}
\toprule
  Condition &    Audio-only &    Video-only &   Audiovisual \\
\midrule
 Audio-only & 0.664 (0.162) & &  \\
 Video-only & 0.237 (0.228) & 0.394 (0.279) &  \\
Audiovisual & 0.663 (0.168) & 0.267 (0.239) & 0.697 (0.165) \\
\bottomrule
\end{tabular}
\end{subtable}

\end{table}

\begin{figure}[hbt!]
\centering
\includegraphics[width=\columnwidth]{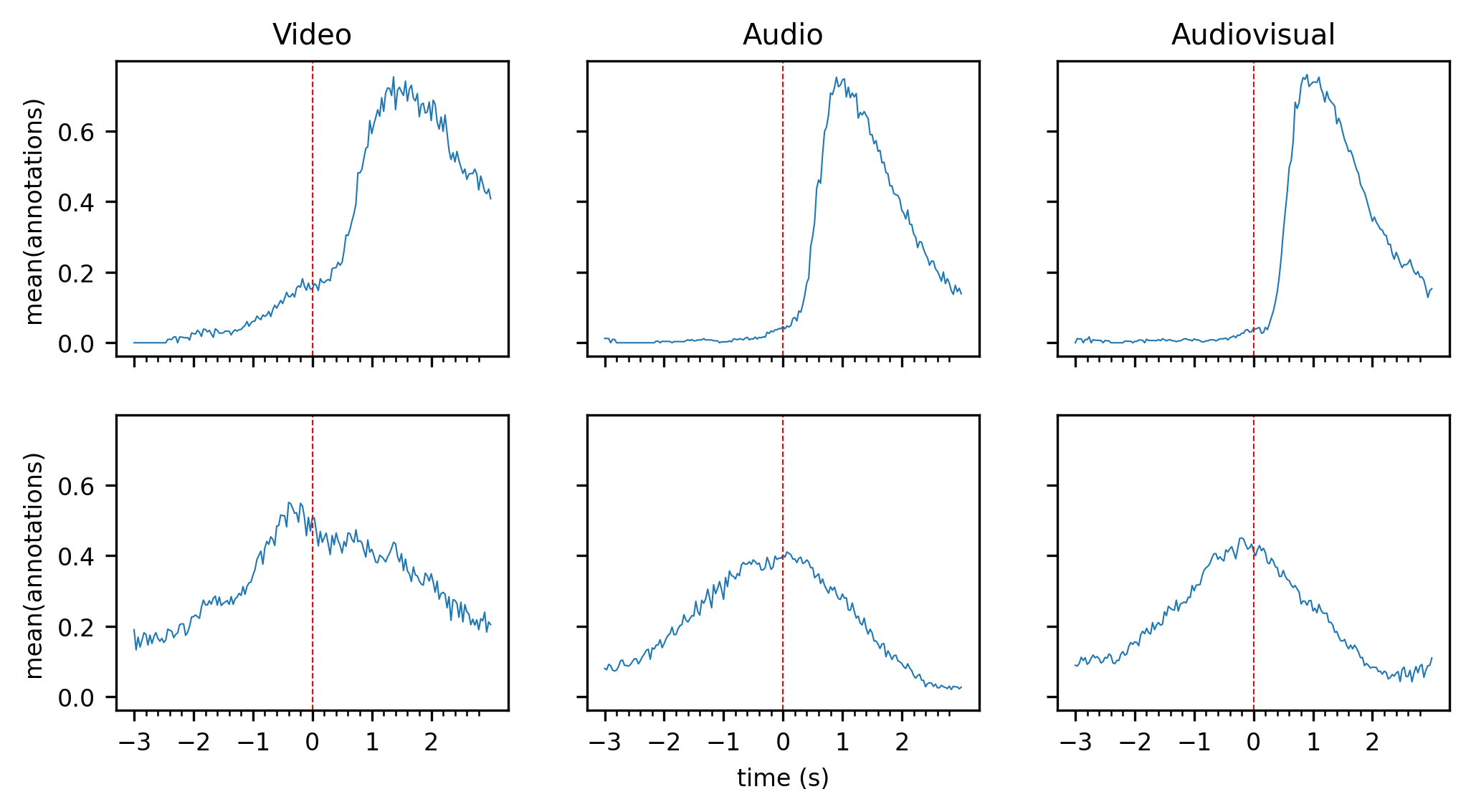}
\caption{Aggregated onsets and offsets w.r.t. reference annotations from different modalities.}
\label{fig:actions:onsets}
\end{figure}

Agreement in intensity estimation (Table \ref{tab:intensity_agreement}) shows a trend consistent with agreement in detection of laughter. The lowest agreements, once again, were found between audio and video ($0.396 \pm 0.186$) and between audiovisual and video conditions ($0.424 \pm 0.183$). These are lower than all within-modality agreement scores, even that of video. This suggests that the \textit{concept} of laughter intensity was perceived differently when audio was available and when it was not. These scores further suggest that audio dominated the audiovisual condition. Note that agreement in laughter intensity was only calculated between examples labeled positively (as laughter) so that these results are independent of detection agreement results (Table \ref{tab:detection_agreement}).

We tested the effect of annotation condition on intensity ratings statistically, making use of a linear mixed effects model with the condition as fixed effect. The annotator ID was used as grouping variable (random effect) to control for annotator-specific variance. We fitted the model only on the subset of positive laughter annotations. We found the condition to have a significant effect on intensity ($p = 0.00223$). 
A cluster bootstrap analysis revealed that laughter was annotated as being significantly less intense in the audiovisual (95\% confidence interval of $[-0.44, -0.0406]$) and video-only conditions (95\% CI of $[-0.45, -0.0482]$). Note that this is a relatively small effect considering the scale of our intensity ratings (1-7).


\begin{figure*}
\centering
\begin{subfigure}{.3\textwidth}
  \centering
  \includegraphics[width=\textwidth]{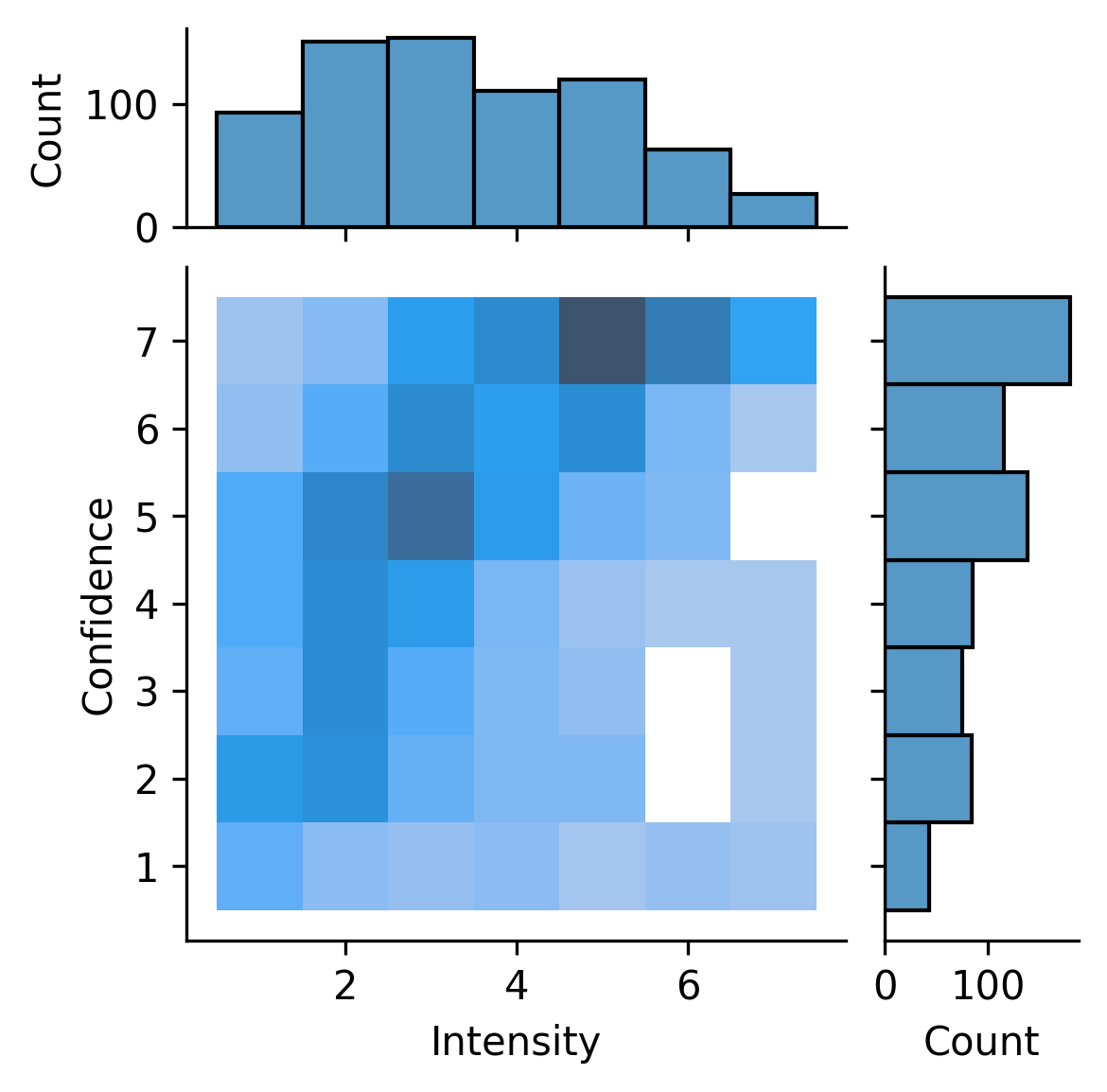}
  \caption{Video condition}
  \label{fig:intensity_confidence:video}
\end{subfigure}%
\begin{subfigure}{.3\textwidth}
  \centering
  \includegraphics[width=\textwidth]{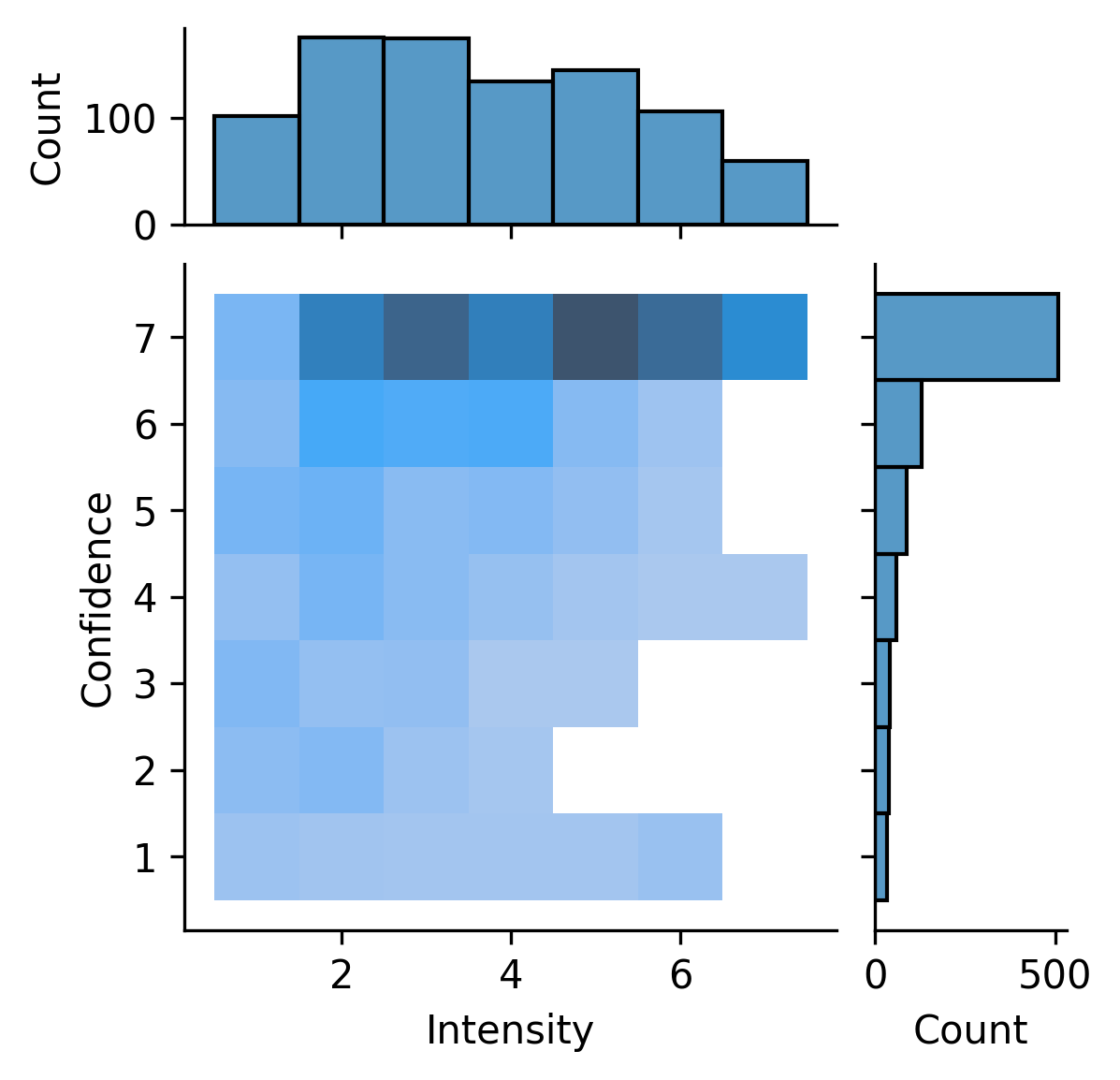}
  \caption{Audio condition}
  \label{fig:intensity_confidence:audio}
\end{subfigure}
\begin{subfigure}{.3\textwidth}
  \centering
  \includegraphics[width=\textwidth]{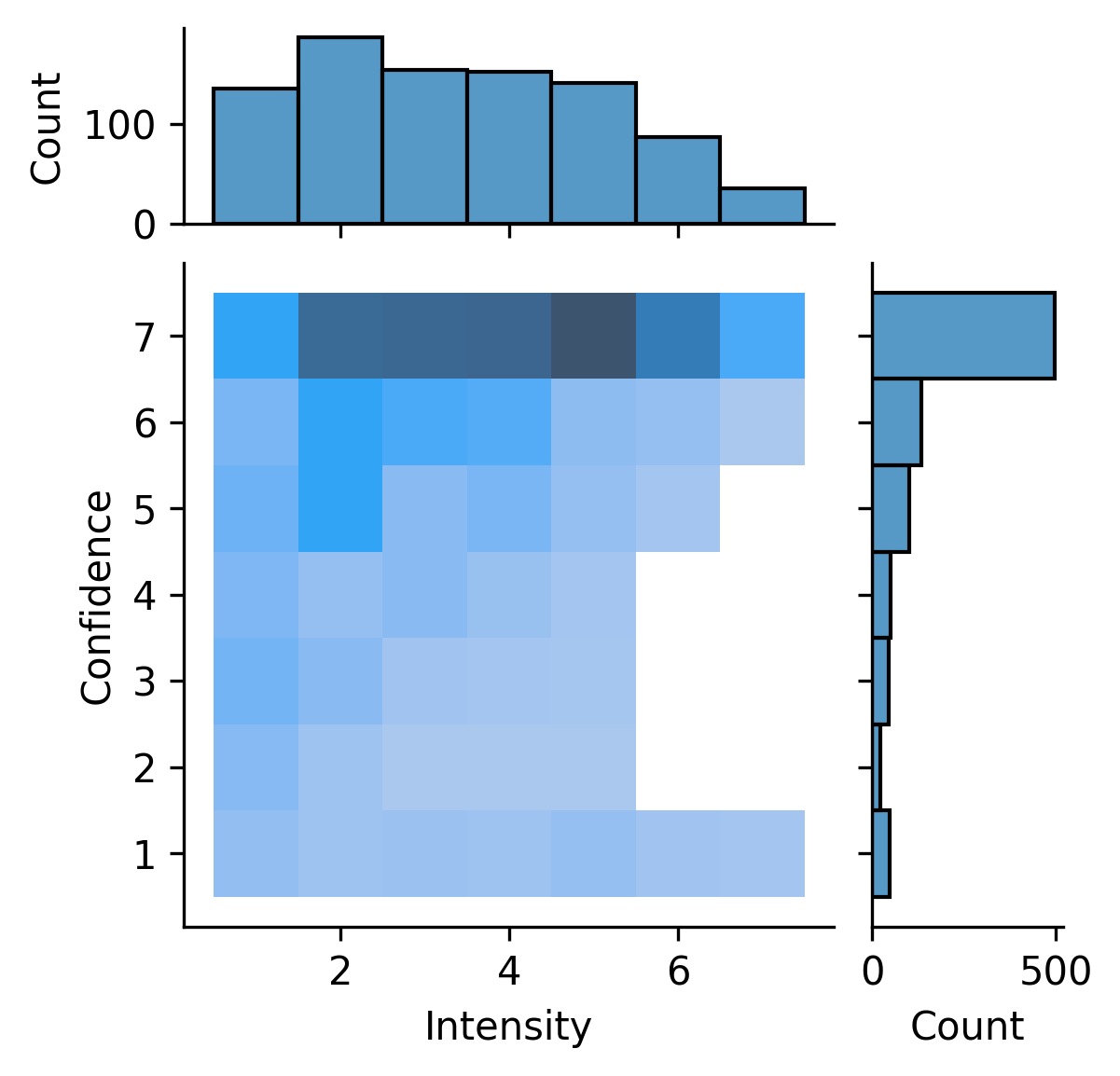}
  \caption{Audiovisual condition}
  \label{fig:intensity_confidence:av}
\end{subfigure}
\caption{Plot showing the joint discrete distribution of confidence and intensity values. Both were annotated using a Likert scale from one to seven. Confidence indicates the confidence of the annotator on their laughter annotation and intensity their estimation of the intensity of the laugh.}
\label{fig:intensity_confidence}
\end{figure*}

To get further clarity about the quality of video-based annotations, we compared them to \textit{reference annotations} from the audiovisual condition. We consider the audiovisual condition to be the most ideal one due to annotators having access to both modalities. However, laughter is not always a clear signal (especially in a dataset like ours) and hence we consider this to be a \textit{reference} set rather than ground truth. We derived a reference set of binary laughter/non-laughter labels via majority voting, for each $(candidate, condition)$ pair, on the annotator ratings (2), and the expert rating (1), for a total of three votes. We used this reference set for calculation of precision and recall scores.

\begin{table}[hbt!]
\centering
\caption{Precision and recall w.r.t. to annotation reference.}
\label{tab:precision_recall}
\begin{tabular}{llll}
\toprule
{} & Audio-only & Video-only & Audiovisual \\
\midrule
Precision &     0.9645 &     0.8915 &      0.9812 \\
Recall    &     0.9405 &     0.7024 &      0.9578 \\
\bottomrule
\end{tabular}
\end{table}

Table \ref{tab:precision_recall} shows the precision and recall scores for the three annotation modalities w.r.t. the reference annotations. Results show that false positives are rare in our annotations. Recall scores show more differences, with video being lower than both audio and audiovisual scores. This aligns with our hypothesis that the video modality is not enough to detect many episodes of laughter (ie. large number of false negatives). As expected, the audiovisual condition had the highest precision and recall. Note however that reference annotation were obtained from audiovisual labels, and this might cause the numbers to be artificially inflated.



Comparing agreement in localization of laughter is less straightforward, since multiple variables are involved.
We decided to do so qualitatively, by plotting the mean value of annotations, across different examples, around reference onsets (rising edge of the binary signal) and offsets (falling edge). Ideally, annotators would agree exactly on the onset of the laugh and we would observe a step-like plot. In practice, onsets and offsets vary per annotation and a curve is observed. Figure \ref{fig:actions:onsets} shows the mean value of annotations around onsets (key pressed) and offsets (key released). These are aggregated over different laughter samples, and show once again better agreement when audio is present. Offsets display less agreement (flatter shape) than onsets. We attribute this to the end of a laugh being often less clear than its start, blending in with speech or other utterances.

We complete our analysis by looking at annotator confidence, as an indication of the difficulty of the task in each modality. Figure \ref{fig:intensity_confidence} we plot the distribution of laughter intensity and confidence values for the three conditions. We used a Likert scale for both of these ratings, and the distributions are therefore discrete. While intensity distributions are similar across the three conditions, the confidence histograms make clear how much more challenging the video-only condition was to annotators. The wider distribution reveals a clear correlation between laughter intensity and confidence in their annotation, as would be expected.

\subsubsection{The role of laughter intensity}
\label{sec:results_intensity}

The results in section \ref{sec:results_agreement} showed that video-only laughter annotations have lower recall than audio-only annotations. We hypothesized, however, that this is likely due to the difficulty of detecting low-intensity laughs, which are likely to have less salient associated body movements.

\begin{figure}[hbt!]
  \centering
  \caption{Laughter recall against intensity of the laughs (as perceived in the audiovisual modality), for video and audio annotations. The $x$ axis corresponds to the middle of percentile bucket (eg. $15$ is the bucket with laughs between the $10th$ and $20th$ percentile). As intensity increases the recall of video-only annotations approaches that of the audio-only annotations}
  \includegraphics[width=0.94\columnwidth]{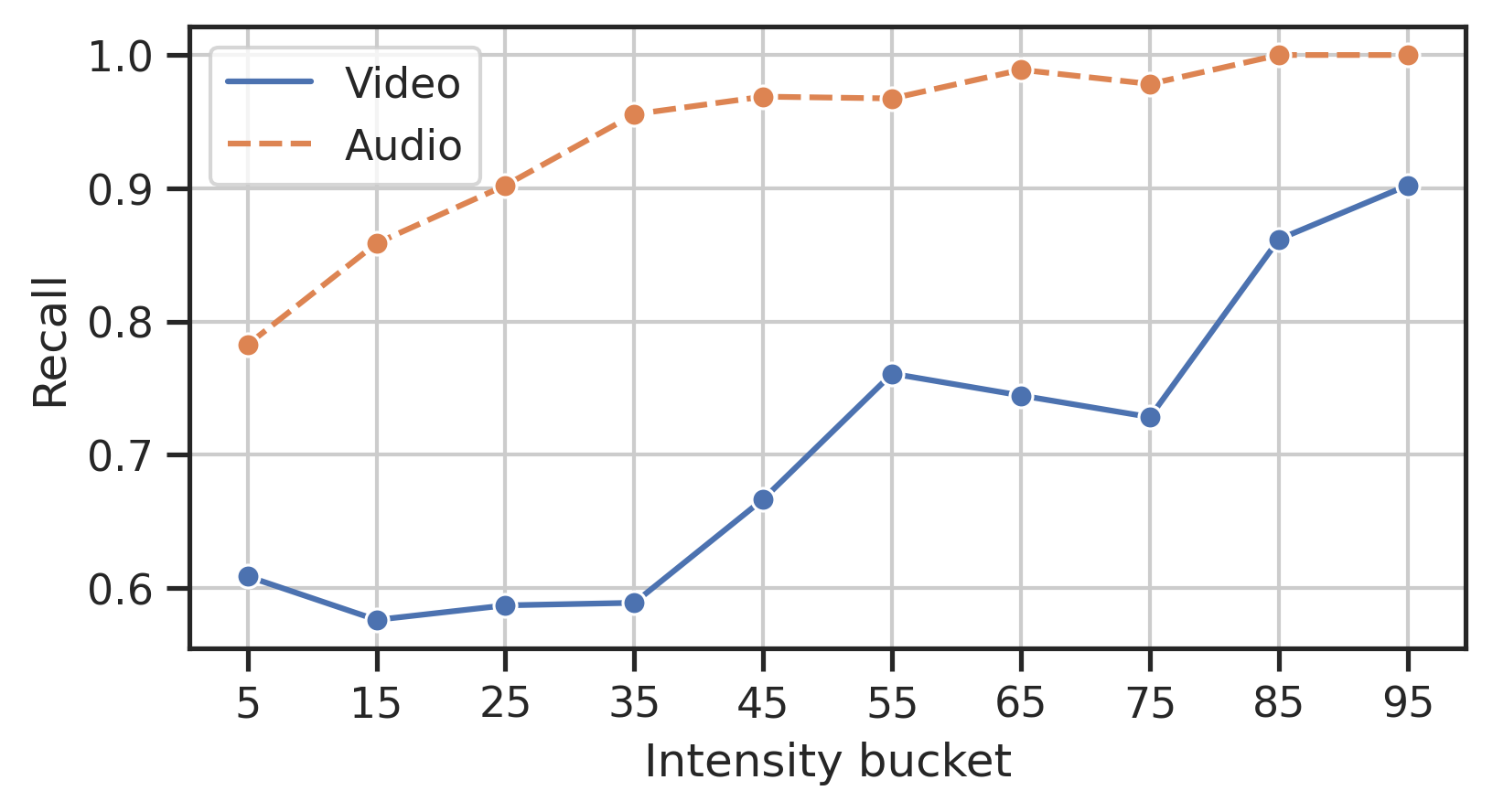}
  \label{fig:recall_intensity}
\end{figure}

To verify this, we separated our dataset by laughter  intensity. We obtained a single consolidated audiovisual intensity rating per example by averaging the intensity ratings from both annotators. We then separated the dataset into 10 intensity buckets, from lowest to highest intensity. To ensure a sufficient number of samples per bucket, we used percentiles to define the bucket sizes, such that bucket $i$ includes laughs between the $(i\times10)th$ and $((i+1)\times10)th$ percentiles of intensity. We computed recall for each bucket. Figure \ref{fig:recall_intensity} plots the results of this analysis. As expected, recall of both audio and video conditions increases with the audiovisual intensity of the laugh. As hypothesized, video recall tends to approach audio recall for the most intense laughs. It stands out, however, that the gap between them never closes completely, even for the $10\%$ most intense laughs. This can be understood in the light of the findings of Section \ref{sec:results_agreement}, were it was shown that intensity ratings in the audio and audiovisual have high agreement, but they both have low agreement with the video-only ratings. Our consolidated audiovisual intensity ratings, therefore, do not reflect intensity as perceived in the video-only condition.

\subsection{Effect of Labeling Modality on Supervised Laughter Tasks}
\label{sec:results_comp}

\begin{figure*}[hbt!]
\centering
\begin{subfigure}{\textwidth}
  \centering
  \caption{Classification into laughter / non-laughter (AUC, higher is better).}
  \includegraphics[width=\textwidth]{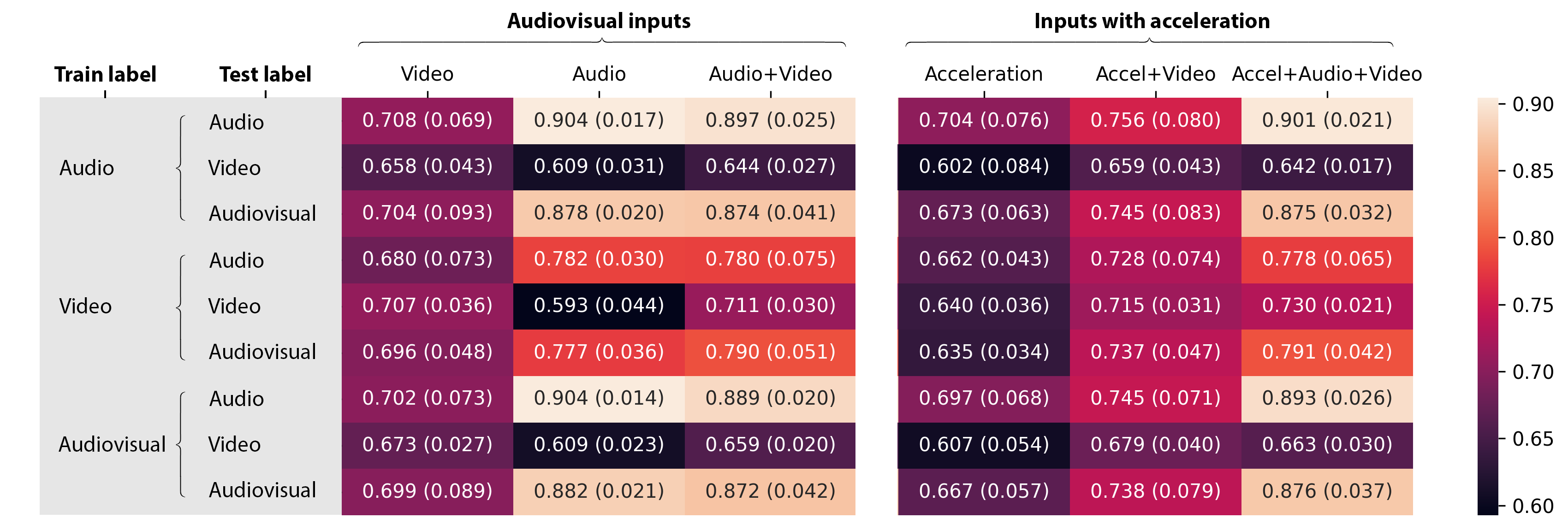}
  \label{fig:results_classification}
\end{subfigure}%

\begin{subfigure}{\textwidth}
  \centering
  \caption{Regression of laughter intensity (MSE, lower is better).}
  \includegraphics[width=\textwidth]{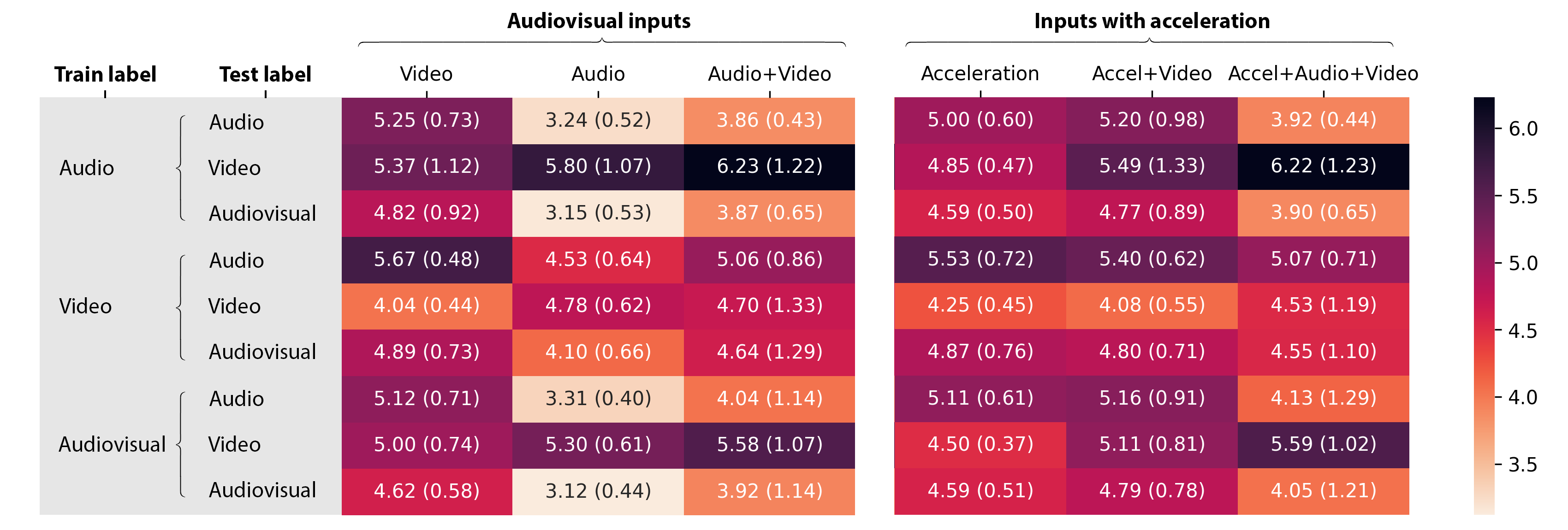}
  \label{fig:results_regression}
\end{subfigure}

\begin{subfigure}{\textwidth}
  \centering
  \caption{Laughter segmentation (AUC, higher is better).}
  \includegraphics[width=\textwidth]{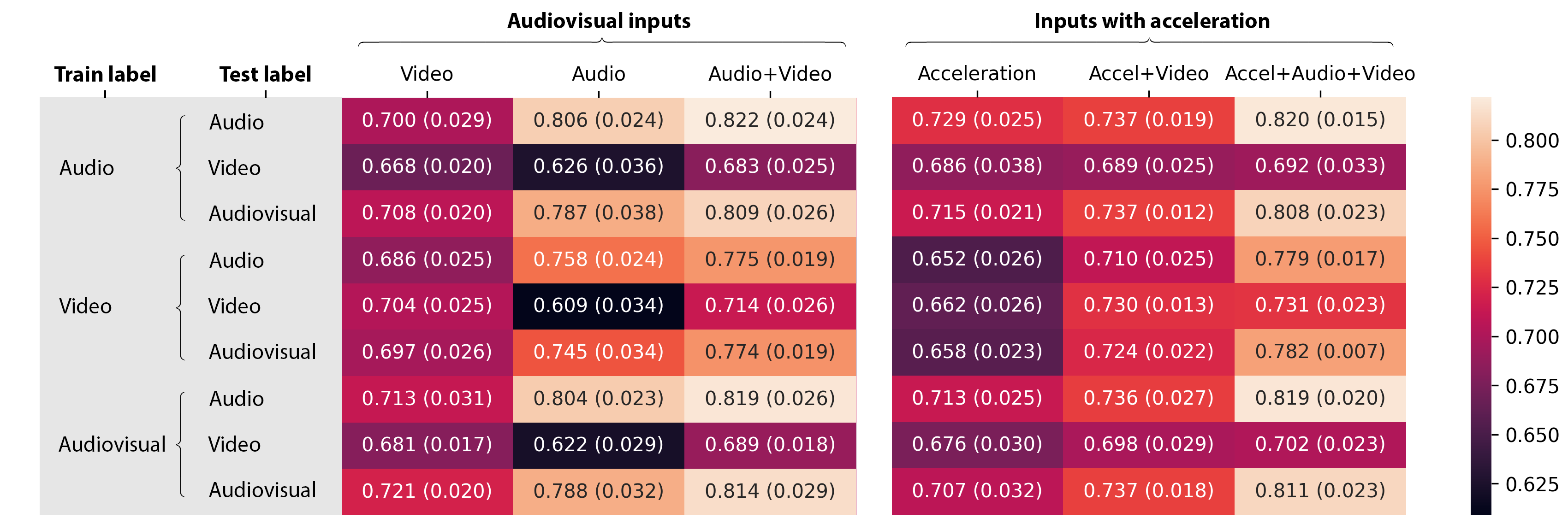}
  \label{fig:results_segmentation}
\end{subfigure}
\caption{Results of our machine learning experiments (10-fold cross-validation). Columns correspond to different model input modalities. Rows correspond to training label modality and testing label modality. For example, $Audio > Video$ indicates a model trained with labels acquired from audio alone, and tested on labels acquired from video alone.}
\label{fig:results}
\end{figure*}

Although the analysis of inter-annotator agreement performed in previous section is relevant to understanding differences in labeling across modalities, it does not ultimately answer the question of how useful annotations acquired from different modalities are for the training of automated models. 

The answer to this question is nuanced. For example, we might have access to body movement annotations of laughter, and want to understand if training a video-based action recognition model with them would help detect actual vocalizations of laughter. However, asking the reverse question is also valid: would audio-based annotations result in a model that detects the characteristic body movements of laughter? Even assuming that we are only interested in detecting the vocalization of laughter, are audio-based labels the most appropriate, or would it be preferable to label from the video modality, for a video-based model?

The goal of this section is to investigate the impact of annotation modality on trained model performance. Machine learning methods can naturally accept different modalities of input data and we are interested in the relationship and possible interactions between input modality, training label modality, and testing label modality.

To this end, in line with the tasks that annotators performed in our human study, we trained and evaluated models for the tasks of laughter detection, intensity estimation and segmentation (Section \ref{sec:method_comp}). For each of these tasks, we evaluated models for all possible combinations of six different input types (acceleration, audio-only, video-only, video+acceleration, audio+video, audiovisual), training label modalities (audio, video, audiovisual) and testing label modalities (audio, video, audiovisual). We used acceleration as an additional input to leverage the wearable data available in our dataset. Wearable acceleration has been found in previous work to be a useful proxy for body movement.

Figure \ref{fig:results_classification} presents the results of our binary laughter / non-laughter classification methods. Positive and negative examples were generated from the human laughter annotations per the procedure in Section \ref{sec:model_windowing}. We evaluated each model using 10-fold cross-validation and the Area under the ROC Curve (AUC) as evaluation metric, as explained in section \ref{sec:evaluation}.

Our classification results reveal several findings. First, it is clear that audio and audiovisual methods, trained on audio or audiovisual labels had the best performances, of up to $0.9$ AUC, except when applied to video-based labels. This is likely explained by these methods detecting many positives that are not labeled in video, due to having low body movement intensity (but which the audio and audiovisual training labels do capture).

In defense of video-based labeling, it stands out that the model with video inputs shows no significant differences across training and testing label modalities. In other words, it did not matter how the labels were acquired for the final performance of the model. The acceleration, and video+acceleration methods had a similar behavior, with no significant differences due to training label modality. This justifies the use of video labels in this setting when they are the easiest to acquire. Furthermore, video labels were even useful for the training of audio methods with an AUC of $~0.78$, a performance drop of less than $0.15$ AUC with respect to the audio labels.

In our regression experiments (Figure \ref{fig:results_regression}) we trained models to estimate laughter intensity ratings. It stands out that, in contrast to classification, for regression most multimodal models performed worse than their audio equivalents for the same labels, meaning that the models did not learn to ignore information from the less informative modalities. Once again we observe the best performances from audio models, trained and tested on audio and audiovisual labels. We also observe that video and acceleration models perform remarkably close to them when trained and tested on video labels, but training on audio and testing on video or vice-versa results in some of the worse performances. This reinforces the findings from the annotation experiments that intensity of laughter in the video and audio modalities do not align.

\section{Discussion}
\label{sec:discussion}

Our annotation agreement results present evidence that annotation of laughter occurrence, intensity and temporal extent can differ substantially across annotation modalities. Per our initial hypothesis, video annotations had lower agreement than audio and audiovisual ones. When comparing against audiovisual reference annotations, we found recall to be worse in the video condition, confirming our hypothesis that video-based annotations are not enough to detect subtle instances of laughter in the wild. Differences in precision scores were lower, with all modalities being close to the 90\% to 95\% range. These findings suggest that video-based annotation of laughter, while feasible, should only be used in applications that require high precision but not high recall. Zooming into the issue of low recall revealed that recall improves for video annotations the more intense the laughs being considered. In the light of previous work \cite{Vernon2016}, this means that laughter annotations done in video in a dataset like ours are more likely to capture humorous laughter, strongly associated to high intensities, than the more common rule-bound conversational laughter.


Regarding differences between audio and audiovisual conditions, our results revealed high within and between-condition agreement ($~0.8$ for detection, $~0.66$ for intensity estimation) between them. These results validate the use of audio as primary modality for laughter annotation, but they are not without nuance. Although they indicate that there was a more clear shared concept being annotated when audio was present, video annotations had higher within-condition agreement than agreement with audio and audiovisual annotations. This suggests that there is a different "concept" being annotated in the video condition with some consistency. Given the low recall of the video condition, we interpret this to indicate that false negatives (w.r.t. audiovisual reference) are missed systematically, likely due to the absence or subtlety of their visual cues. Systematic false positives across annotators also likely contribute to these results, though to a smaller degree. In other words, there appears to be incongruence in the perception of laughter occurrence. 

These results set the stage for the question explored in our machine learning analysis: is perception of laughter in the visual modality a meaningful concept to annotate for the purpose of building detectors, despite it's incongruence with audiovisual laughter? 

Importantly, we measured a similar incongruence in laughter intensity ratings, where only positively-labeled segments were included in the agreement calculations, indicating that laughter intensity is not perceived in the same way when audio is present and when it is not. Such incongruences in laughter intensity across modalities have only been studied in the context of laughter synthesis. Niewiadomski et al. found that laughter episodes with incongruent body movement and vocalization intensities were rated as less believable \cite{Niewiadomski2015}. This would seem to go against our results, which suggest that significant incongruence exists in in-the-wild laughter perception. However, the magnitude  of the incongruencies used (which can be controlled in a synthesis study, but not in the wild) could explain this discrepancy.

Our results have implications in studies of laughter intensity \cite{Mancini2012, Niewiadomski2012, McKeown2013, DiLascio2019, Niewiadomski2015, Haddad, Curran2018}, suggesting that the concept of laughter intensity should not be treated as a scalar property of the laughter episode, but rather as a nuanced evaluation affected especially by the modalities available to the observer. In particular, the question of whether a clear distinction should be made between the intensity of body movements and the intensity of the sound of laughter deserves consideration. McKeown et al. already asked the question of whether laughter body movement intensity itself should be considered multi-dimensional \cite{McKeown2013} (which we also consider relevant to ask), but the distinction between visual and auditory intensity has not been considered before, to the best of our knowledge.

Our findings lead us to the fundamental question of what is laughter intensity in the wild. Would our results translate to real-life evaluation of laughter intensity, as opposed to via recordings? While in our dataset subjects prioritized audio in the multimodal condition, it is not clear if body movement information would be prioritized in other datasets in which it is easier to perceive (ie. with consistent access to the face or upper body), or in which the audio is harder to perceive. We consider it likely that in such cases visual information will at least play a more important role, but more work is necessary to have an answer to these questions.

Despite the lower inter-annotator agreement in the video conditions, our machine learning experiments with different combinations of model inputs, training label modalities, and testing label modalities, revealed that model performance was the same across labels for models trained using video and acceleration inputs, both of which capture body movements. This was regardless of the evaluation modality. In other words, annotating laughter (traditionally understood primarily as a vocalization) from the video modality only may be perfectly valid when the goal is to optimize model performance. We think that the reason for such results is explained by our human annotation analysis. Concretely, episodes with lower intensity were most commonly missed (w.r.t. to the audiovisual reference). The subtlety of these training samples would presumably make them more challenging for the learning algorithm, and therefore their absence would not have an adverse effect on performance. We obtained these results in a challenging dataset, where many positive (audiovisual) laughter episodes were missed by annotators, and using a state-of-the-art action recognition 3D-CNN. It is once again unclear whether these results would translate to a dataset with more consistent access to, for example, facial visual information. The presence of visual cues could improve the model, but their subtlety could be a challenge to most state-of-the-art models.

Our results provide validation for previous works using video-only labelling to train laughter assessment models from body movements \cite{Mancini2012, Cu2017}, and datasets providing video-only annotations \cite{Cabrera-Quiros2018a}. However, it should be noted that model performance may not be the only relevant variable in analysis of such models. The fact that annotations obtained from video are largely incongruent with audiovisual annotations should be a consideration in their design. 

We think that these results could have wider implications if they generalize to other multi-modal social signals with manifestations in body movement. Speaking status (or voice activity) and back-channels have been of interest in previous work \cite{Truong2011, Beyan2020}. Video-only annotations of speaking status have been used in previous work \cite{Cabrera-Quiros2018a, Gedik2017, Raman2022b}, but the implications in model performance of this annotation choice have not been explored. Our results would suggest that it is possible to annotate speaking from video alone without an adverse effect on the model's ability to detect speech, but further work is necessary to provide validation for other multimodal social signals besides laughter.  

\subsection{Limitations}

We consider the main limitation of our work to be that we used only one dataset in our experiments. Our dataset is however representative of one of the most challenging scenarios for perception of laughter from video: with little access to the face of the participants, different views and distances to camera, low light conditions, and significant occlusion of parts of the body from other participants in the scene. We therefore considered it a useful data point to study. We expect that more traditional front-facing datasets with consistent access to the body and face of the subjects will result in lower differences in agreement and model performance between the video condition and the audio and audiovisual ones. We think it is possible that clear access to the face will negate the incongruence observed in laughter intensity ratings, since facial features may share more information with the laughter vocalization than overall body movement does.

\section*{Acknowledgements}
This research is supported by the Netherlands Organization for Scientific Research (NWO) under project number 639.022.606.

\ifCLASSOPTIONcaptionsoff
  \newpage
\fi



\bibliographystyle{IEEEtran}

\bibliography{library.bib}

\begin{thebibliography}{10}
\providecommand{\url}[1]{#1}
\csname url@samestyle\endcsname
\providecommand{\newblock}{\relax}
\providecommand{\bibinfo}[2]{#2}
\providecommand{\BIBentrySTDinterwordspacing}{\spaceskip=0pt\relax}
\providecommand{\BIBentryALTinterwordstretchfactor}{4}
\providecommand{\BIBentryALTinterwordspacing}{\spaceskip=\fontdimen2\font plus
\BIBentryALTinterwordstretchfactor\fontdimen3\font minus
  \fontdimen4\font\relax}
\providecommand{\BIBforeignlanguage}[2]{{%
\expandafter\ifx\csname l@#1\endcsname\relax
\typeout{** WARNING: IEEEtran.bst: No hyphenation pattern has been}%
\typeout{** loaded for the language `#1'. Using the pattern for}%
\typeout{** the default language instead.}%
\else
\language=\csname l@#1\endcsname
\fi
#2}}
\providecommand{\BIBdecl}{\relax}
\BIBdecl

\bibitem{Darwin1887}
C.~Darwin, ``Expression of the emotions in {Man} and {Animals},''
  \emph{Nature}, vol.~36, no. 926, pp. 294--295, 1887, iSBN: 0195158067.

\bibitem{Burgoon}
J.~K. Burgoon, N.~Magnenat-Thalmann, M.~Pantic, and A.~Vinciarelli,
  \emph{Social {Signal} {Processing}}, 2017.

\bibitem{Vinciarelli2009}
\BIBentryALTinterwordspacing
A.~Vinciarelli, M.~Pantic, and H.~Bourlard, ``Social signal processing:
  {Survey} of an emerging domain,'' \emph{Image and Vision Computing}, vol.~27,
  no.~12, pp. 1743--1759, 2009, publisher: Elsevier B.V. [Online]. Available:
  \url{http://dx.doi.org/10.1016/j.imavis.2008.11.007}
\BIBentrySTDinterwordspacing

\bibitem{Mancini2012}
M.~Mancini, G.~Varni, D.~Glowinski, and G.~Volpe, ``Computing and evaluating
  the body laughter index,'' \emph{Lecture Notes in Computer Science (including
  subseries Lecture Notes in Artificial Intelligence and Lecture Notes in
  Bioinformatics)}, vol. 7559 LNCS, pp. 90--98, 2012, iSBN: 9783642340130.

\bibitem{Niewiadomski2012}
R.~Niewiadomski, J.~Urbain, C.~Pelachaud, and T.~Dutoit, ``Finding out the
  audio and visual features that influence the perception of laughter intensity
  and differ in inhalation and exhalation phases,'' \emph{Proceedings of the
  4th Interna- tionalWorkshop onCorpora forResearch on Emotion}, pp. 25--32,
  2012.

\bibitem{McKeown2013}
G.~McKeown, W.~Curran, D.~Kane, R.~McCahon, H.~J. Griffin, C.~McLoughlin, and
  N.~Bianchi-Berthouze, ``Human perception of laughter from context-free whole
  body motion dynamic stimuli,'' \emph{Proceedings - 2013 Humaine Association
  Conference on Affective Computing and Intelligent Interaction, ACII 2013},
  pp. 306--311, 2013, publisher: IEEE ISBN: 9780769550480.

\bibitem{DiLascio2019}
\BIBentryALTinterwordspacing
E.~Di~Lascio, S.~Gashi, and S.~Santini, ``Laughter {Recognition} {Using}
  {Non}-invasive {Wearable} {Devices},'' 2019, iSBN: 9781450361262. [Online].
  Available: \url{https://doi.org/10.1145/3329189.3329216}
\BIBentrySTDinterwordspacing

\bibitem{Niewiadomski2015}
R.~Niewiadomski, Y.~Ding, M.~Mancini, C.~Pelachaud, G.~Volpe, and A.~Camurri,
  ``Perception of intensity incongruence in synthesized multimodal expressions
  of laughter,'' \emph{2015 International Conference on Affective Computing and
  Intelligent Interaction, ACII 2015}, pp. 684--690, 2015, publisher: IEEE
  ISBN: 9781479999538.

\bibitem{Haddad}
K.~E. Haddad, S.~N. Chakravarthula, and J.~Kennedy, ``Smile and {Laugh}
  {Dynamics} in {Naturalistic} {Dyadic} {Interactions} : {Intensity} {Levels},
  {Sequences} and {Roles},'' iSBN: 9781450368605.

\bibitem{Curran2018}
W.~Curran, G.~J. McKeown, M.~Rychlowska, E.~André, J.~Wagner, and
  F.~Lingenfelser, ``Social context disambiguates the interpretation of
  laughter,'' \emph{Frontiers in Psychology}, vol.~8, no. JAN, pp. 1--12, 2018.

\bibitem{Mazzocconi2020}
C.~Mazzocconi, Y.~Tian, and J.~Ginzburg, ``What's your laughter doing there?
  {A} taxonomy of the pragmatic functions of laughter,'' \emph{IEEE
  Transactions on Affective Computing}, pp. 1--1, 2020, conference Name: IEEE
  Transactions on Affective Computing.

\bibitem{Petridis2008a}
\BIBentryALTinterwordspacing
S.~Petridis and M.~Pantic, ``Audiovisual discrimination between laughter and
  speech,'' \emph{2008 IEEE International Conference on Acoustics, Speech and
  Signal Processing}, pp. 5117--5120, 2008, iSBN: 978-1-4244-1483-3. [Online].
  Available: \url{http://ieeexplore.ieee.org/document/4518810/}
\BIBentrySTDinterwordspacing

\bibitem{Petridis2013}
\BIBentryALTinterwordspacing
S.~Petridis, B.~Martinez, and M.~Pantic, ``The {MAHNOB} {Laughter} database,''
  \emph{Image and Vision Computing}, vol.~31, no.~2, pp. 186--202, 2013,
  publisher: Elsevier B.V. ISBN: 0262-8856. [Online]. Available:
  \url{http://dx.doi.org/10.1016/j.imavis.2012.08.014}
\BIBentrySTDinterwordspacing

\bibitem{Truong2007}
K.~P. Truong and D.~A. van Leeuwen, ``Automatic discrimination between laughter
  and speech,'' \emph{Speech Communication}, vol.~49, no.~2, pp. 144--158,
  2007, iSBN: 0167-6393.

\bibitem{Truong2012}
K.~P. Truong and J.~Trouvain, ``On the acoustics of overlapping laughter in
  conversational speech,'' \emph{13th Annual Conference of the International
  Speech Communication Association 2012, INTERSPEECH 2012}, vol.~1, pp.
  850--853, 2012, iSBN: 9781622767595.

\bibitem{Cabrera-Quiros2018a}
L.~Cabrera-Quiros, A.~Demetriou, E.~Gedik, L.~van~der Meij, and H.~Hung, ``The
  {MatchNMingle} {Dataset}: {A} {Novel} {Multi}-{Sensor} {Resource} for the
  {Analysis} of {Social} {Interactions} and {Group} {Dynamics} {In}-the-{Wild}
  {During} {Free}-{Standing} {Conversations} and {Speed} {Dates},'' \emph{IEEE
  Transactions on Affective Computing}, vol.~12, no.~1, pp. 113--130, Jan.
  2021, conference Name: IEEE Transactions on Affective Computing.

\bibitem{Gedik2017}
E.~Gedik and H.~Hung, ``Personalised models for speech detection from body
  movements using transductive parameter transfer,'' \emph{Personal and
  Ubiquitous Computing}, vol.~21, no.~4, pp. 723--737, 2017, publisher:
  Springer London ISBN: 9781450344623.

\bibitem{Vargas2019a}
J.~Vargas and H.~Hung, ``{CNNs} and {Fisher} {Vectors} for {No}-{Audio}
  {Multimodal} {Speech} {Detection},'' pp. 11--13, 2019.

\bibitem{Raman2022b}
\BIBentryALTinterwordspacing
C.~Raman, J.~Vargas-Quiros, S.~Tan, E.~Gedik, A.~Islam, and H.~Hung,
  ``{ConfLab}: {A} {Rich} {Multimodal} {Multisensor} {Dataset} of
  {Free}-{Standing} {Social} {Interactions} in the {Wild},'' Jul. 2022,
  arXiv:2205.05177 [cs]. [Online]. Available:
  \url{http://arxiv.org/abs/2205.05177}
\BIBentrySTDinterwordspacing

\bibitem{Ginzburg2015}
J.~Ginzburg, E.~Breitholtz, R.~Cooper, J.~Hough, and T.~Ye,
  ``\BIBforeignlanguage{en}{Understanding {Laughter}},''
  \emph{\BIBforeignlanguage{en}{20th Amsterdam Colloquium}}, p.~11, 2015.

\bibitem{Oatley2014}
\BIBentryALTinterwordspacing
K.~Oatley and P.~Johnson-Laird, ``\BIBforeignlanguage{en}{Cognitive approaches
  to emotions},'' \emph{\BIBforeignlanguage{en}{Trends in Cognitive Sciences}},
  vol.~18, no.~3, pp. 134--140, Mar. 2014. [Online]. Available:
  \url{https://linkinghub.elsevier.com/retrieve/pii/S1364661313002866}
\BIBentrySTDinterwordspacing

\bibitem{Scherer2009a}
\BIBentryALTinterwordspacing
K.~R. Scherer, ``The dynamic architecture of emotion: {Evidence} for the
  component process model,'' \emph{Cognition and Emotion}, vol.~23, no.~7, pp.
  1307--1351, Nov. 2009, publisher: Routledge \_eprint:
  https://doi.org/10.1080/02699930902928969. [Online]. Available:
  \url{https://doi.org/10.1080/02699930902928969}
\BIBentrySTDinterwordspacing

\bibitem{Gervais2005}
\BIBentryALTinterwordspacing
M.~Gervais and D.~S. Wilson, ``{THE} {EVOLUTION} {AND} {FUNCTIONS} {OF}
  {LAUGHTER} {AND} {HUMOR}: {A} {SYNTHETIC} {APPROACH},'' vol.~80, no.~4, pp.
  1--38, 2005. [Online]. Available:
  \url{papers://061a454d-2726-4f93-8196-167c101a1d0a/Paper/p3275}
\BIBentrySTDinterwordspacing

\bibitem{Miller2009}
\BIBentryALTinterwordspacing
M.~Miller and W.~F. Fry, ``\BIBforeignlanguage{en}{The effect of mirthful
  laughter on the human cardiovascular system},''
  \emph{\BIBforeignlanguage{en}{Medical Hypotheses}}, vol.~73, no.~5, pp.
  636--639, Nov. 2009. [Online]. Available:
  \url{https://linkinghub.elsevier.com/retrieve/pii/S0306987709002898}
\BIBentrySTDinterwordspacing

\bibitem{Gillick2021}
\BIBentryALTinterwordspacing
J.~Gillick, W.~Deng, K.~Ryokai, and D.~Bamman, ``\BIBforeignlanguage{en}{Robust
  {Laughter} {Detection} in {Noisy} {Environments}},'' in
  \emph{\BIBforeignlanguage{en}{Interspeech 2021}}.\hskip 1em plus 0.5em minus
  0.4em\relax ISCA, Aug. 2021, pp. 2481--2485. [Online]. Available:
  \url{https://www.isca-speech.org/archive/interspeech_2021/gillick21_interspeech.html}
\BIBentrySTDinterwordspacing

\bibitem{Petridis2013a}
S.~Petridis, M.~Leveque, and M.~Pantic, ``Audiovisual detection of laughter in
  human-machine interaction,'' \emph{Proceedings - 2013 Humaine Association
  Conference on Affective Computing and Intelligent Interaction, ACII 2013},
  pp. 129--134, 2013, publisher: IEEE ISBN: 9780769550480.

\bibitem{Griffin2013}
H.~J. Griffin, M.~S. Aung, B.~Romera-Paredes, C.~McLoughlin, G.~McKeown,
  W.~Curran, and N.~Bianchi-Berthouze, ``Laughter type recognition from whole
  body motion,'' \emph{Proceedings - 2013 Humaine Association Conference on
  Affective Computing and Intelligent Interaction, ACII 2013}, no. September,
  pp. 349--355, 2013, iSBN: 9780769550480.

\bibitem{Griffin2015}
------, ``Perception and automatic recognition of laughter from whole-body
  motion: {Continuous} and categorical perspectives,'' \emph{IEEE Transactions
  on Affective Computing}, vol.~6, no.~2, pp. 165--178, 2015, iSBN: 1949-3045
  VO - 6.

\bibitem{Silvervarg2012}
\BIBentryALTinterwordspacing
R.~Niewiadomski and C.~Pelachaud, ``Towards {Multimodal} {Expression} of
  {Laughter},'' vol. 7502, no.~0, p. 6221, 2012, iSBN: 978-3-642-33196-1.
  [Online]. Available: \url{http://link.springer.com/10.1007/978-3-642-33197-8}
\BIBentrySTDinterwordspacing

\bibitem{Petridis2015}
S.~Petridis, ``A {Short} {Introduction} to {Laughter},'' pp. 1--24, 2015.

\bibitem{McKeown2015b}
G.~McKeown, W.~Curran, J.~Wagner, F.~Lingenfelser, and E.~André, ``The
  {Belfast} storytelling database: {A} spontaneous social interaction database
  with laughter focused annotation,'' in \emph{2015 {International}
  {Conference} on {Affective} {Computing} and {Intelligent} {Interaction}
  ({ACII})}, Sep. 2015, pp. 166--172, iSSN: 2156-8111.

\bibitem{Jansen2020}
M.-p. Jansen, K.~P. Truong, D.~S. Nazareth, and D.~K.~J. Heylen, ``Introducing
  {MULAI} : {A} {Multimodal} {Database} of {Laughter} during {Dyadic}
  {Interactions},'' no. May, pp. 4333--4342, 2020.

\bibitem{Provine2000}
R.~R. C. N. . . B. L. D. S. C. m. B. L. H.~Y. Provine, ``Laughter : a
  scientific investigation,'' no. March, 2000, iSBN: 0571191894 (pbk) :
  ¹12.99.

\bibitem{Truong2019}
K.~P. Truong, J.~Trouvain, and M.-p. Jansen, ``Towards an annotation scheme for
  complex laughter in speech corpora,'' pp. 529--533, 2019.

\bibitem{ElHaddad2018}
K.~El~Haddad, H.~Cakmak, and T.~Dutoit, ``On {Laughter} {Intensity} {Level}:
  {Analysis} and {Estimation},'' Sep. 2018.

\bibitem{Vernon2016}
\BIBentryALTinterwordspacing
S.~Dupont, H.~Çakmak, W.~Curran, T.~Dutoit, J.~Hofmann, G.~McKeown,
  O.~Pietquin, T.~Platt, W.~Ruch, and J.~Urbain, \emph{Laughter {Research}: {A}
  {Review} of the {ILHAIRE} {Project}}, 2016, vol. 105. [Online]. Available:
  \url{http://link.springer.com/10.1007/978-3-319-31056-5}
\BIBentrySTDinterwordspacing

\bibitem{Provine1993}
R.~R. Provine, ``Laughter {Punctuates} {Speech}: {Linguistic}, {Social} and
  {Gender} {Contexts} of {Laughter},'' \emph{Ethology}, vol.~95, no.~4, pp.
  291--298, 1993, iSBN: 0179-1613.

\bibitem{Holt2010}
\BIBentryALTinterwordspacing
E.~Holt, ``The last laugh : {Shared} laughter and topic termination,''
  \emph{Journal of Pragmatics}, vol.~42, no.~6, pp. 1513--1525, 2010,
  publisher: Elsevier B.V. [Online]. Available:
  \url{http://dx.doi.org/10.1016/j.pragma.2010.01.011}
\BIBentrySTDinterwordspacing

\bibitem{Odonnell-Trujillo1983}
N.~O’donnell-Trujillo and K.~Adams, ``Heheh in conversation: {Some}
  coordinating accomplishments of laughter,'' \emph{Western Journal of Speech
  Communication}, vol.~47, no.~2, pp. 175--191, 1983.

\bibitem{Poyatos1993}
\BIBentryALTinterwordspacing
F.~Poyatos, ``\BIBforeignlanguage{en}{The many voices of laughter: {A} new
  audible-visual paralinguistic approach},'' vol.~93, no. 1-2, pp. 61--82, Jan.
  1993, publisher: De Gruyter Mouton Section: Semiotica. [Online]. Available:
  \url{http://www.degruyter.com/document/doi/10.1515/semi.1993.93.1-2.61/pdf}
\BIBentrySTDinterwordspacing

\bibitem{Truong2005}
K.~Truong and D.~V. Leeuwen, ``Automatic {Detection} of {Laughter},''
  \emph{Interspeech}, pp. 485--488, 2005, iSBN: 1855212986.

\bibitem{Petridis2008}
S.~Petridis and M.~Pantic, ``Audiovisual {Laughter} {Detection} {Based} on
  {Temporal} {Features},'' \emph{Belgian/Netherlands Artificial Intelligence
  Conference}, pp. 351--352, 2008, iSBN: 9781605581989.

\bibitem{Petridis2011}
------, ``Audiovisual {Discrimination} {Between} {Speech} and {Laughter}: {Why}
  and {When} {Visual} {Information} {Might} {Help},'' \emph{Why and When
  Visual}, vol.~13, no.~2, pp. 216--234, 2011.

\bibitem{Beyan2020}
C.~Beyan, M.~Shahid, and V.~Murino, ``{RealVAD}: {A} {Real}-world {Dataset} and
  {A} {Method} for {Voice} {Activity} {Detection} by {Body} {Motion}
  {Analysis},'' \emph{x}, vol. 9210, no.~c, pp. 1--16, 2020.

\bibitem{Niewiadomski2016}
R.~Niewiadomski, M.~Mancini, G.~Varni, G.~Volpe, and A.~Camurri, ``Automated
  {Laughter} {Detection} from {Full}-{Body} {Movements},'' \emph{IEEE
  Transactions on Human-Machine Systems}, vol.~46, no.~1, pp. 113--123, 2016.

\bibitem{Cu2017}
\BIBentryALTinterwordspacing
J.~Cu, M.~B. Luz, M.~Nocum, and T.~J. Purganan, ``Affective {Laughter}
  {Expressions} from {Body} {Movements},'' vol. 10004, pp. 139--145, 2017,
  iSBN: 978-3-319-60674-3. [Online]. Available:
  \url{http://link.springer.com/10.1007/978-3-319-60675-0}
\BIBentrySTDinterwordspacing

\bibitem{Haddad2015}
K.~E. Haddad, S.~Dupont, J.~Urbain, and T.~Dutoit, ``Speech-laughs: {An}
  {HMM}-based approach for amused speech synthesis,'' in \emph{2015 {IEEE}
  {International} {Conference} on {Acoustics}, {Speech} and {Signal}
  {Processing} ({ICASSP})}, Apr. 2015, pp. 4939--4943, iSSN: 2379-190X.

\bibitem{Trouvain2003}
J.~Trouvain, ``Segmenting phonetic units in laughter,'' \emph{Proc. ICPhS '03},
  pp. 2793--2796, 2003, iSBN: 1876346485.

\bibitem{Jordan2010}
T.~R. Jordan and L.~Abedipour, ``The importance of laughing in your face:
  {Influences} of visual laughter on auditory laughter perception,''
  \emph{Perception}, vol.~39, no.~9, pp. 1283--1285, 2010, iSBN:
  0301-0066{\textbackslash}n1468-4233.

\bibitem{Carletta2007}
J.~Carletta, ``Unleashing the killer corpus: {Experiences} in creating the
  multi-everything {AMI} {Meeting} {Corpus},'' \emph{Language Resources and
  Evaluation}, vol.~41, no.~2, pp. 181--190, 2007.

\bibitem{McKeown2012}
G.~McKeown, M.~Valstar, R.~Cowie, M.~Pantic, and M.~Schröder, ``The {SEMAINE}
  database: {Annotated} multimodal records of emotionally colored conversations
  between a person and a limited agent,'' \emph{IEEE Transactions on Affective
  Computing}, vol.~3, no.~1, pp. 5--17, 2012, iSBN: 1949-3045.

\bibitem{Reuderink2008}
B.~Reuderink, M.~Poel, K.~Truong, R.~Poppe, and M.~Pantic, ``Decision-level
  fusion for audio-visual laughter detection,'' \emph{Lecture Notes in Computer
  Science (including subseries Lecture Notes in Artificial Intelligence and
  Lecture Notes in Bioinformatics)}, vol. 5237 LNCS, pp. 137--148, 2008, iSBN:
  3540858520.

\bibitem{Mancini2013}
M.~Mancini, J.~Hofmann, T.~Platt, G.~Volpe, G.~Varni, D.~Glowinski, W.~Ruch,
  and A.~Camurri, ``Towards automated full body detection of laughter driven by
  human expert annotation,'' \emph{Proceedings - 2013 Humaine Association
  Conference on Affective Computing and Intelligent Interaction, ACII 2013},
  pp. 757--762, 2013, publisher: IEEE ISBN: 9780769550480.

\bibitem{Alameda-Pineda2015}
\BIBentryALTinterwordspacing
X.~Alameda-Pineda, J.~Staiano, R.~Subramanian, L.~Batrinca, E.~Ricci, B.~Lepri,
  O.~Lanz, and N.~Sebe, ``{SALSA}: {A} {Novel} {Dataset} for {Multimodal}
  {Group} {Behavior} {Analysis},'' Jun. 2015, arXiv:1506.06882 [cs]. [Online].
  Available: \url{http://arxiv.org/abs/1506.06882}
\BIBentrySTDinterwordspacing

\bibitem{Tan2020}
\BIBentryALTinterwordspacing
Z.~H. Tan, A.~k. Sarkar, and N.~Dehak, ``{rVAD}: {An} unsupervised
  segment-based robust voice activity detection method,'' \emph{Computer Speech
  and Language}, vol.~59, pp. 1--21, 2020, arXiv: 1906.03588 Publisher:
  Elsevier Ltd. [Online]. Available:
  \url{https://doi.org/10.1016/j.csl.2019.06.005}
\BIBentrySTDinterwordspacing

\bibitem{Crystal1990}
\BIBentryALTinterwordspacing
T.~H. Crystal and A.~S. House, ``\BIBforeignlanguage{en}{Articulation rate and
  the duration of syllables and stress groups in connected speech},''
  \emph{\BIBforeignlanguage{en}{The Journal of the Acoustical Society of
  America}}, vol.~88, no.~1, pp. 101--112, Jul. 1990. [Online]. Available:
  \url{http://asa.scitation.org/doi/10.1121/1.399955}
\BIBentrySTDinterwordspacing

\bibitem{Elan2021}
\BIBentryALTinterwordspacing
M.~P.~I. for Psycholinguistics, ``{ELAN} [{Computer} software].'' Nijmegen, The
  Netherlands, 2021. [Online]. Available: \url{https://archive.mpi.nl/tla/elan}
\BIBentrySTDinterwordspacing

\bibitem{Quiros2022}
\BIBentryALTinterwordspacing
J.~V. Quiros, S.~Tan, C.~Raman, L.~Cabrera-Quiros, and H.~Hung,
  ``\BIBforeignlanguage{en}{Covfee: an extensible web framework for
  continuous-time annotation of human behavior},'' in
  \emph{\BIBforeignlanguage{en}{Understanding {Social} {Behavior} in {Dyadic}
  and {Small} {Group} {Interactions}}}.\hskip 1em plus 0.5em minus 0.4em\relax
  PMLR, Mar. 2022, pp. 265--293, iSSN: 2640-3498. [Online]. Available:
  \url{https://proceedings.mlr.press/v173/vargas-quiros22a.html}
\BIBentrySTDinterwordspacing

\bibitem{prolific}
\BIBentryALTinterwordspacing
``Prolific,'' London, UK, 2014. [Online]. Available:
  \url{https://www.prolific.co}
\BIBentrySTDinterwordspacing

\bibitem{Mariooryad2015}
S.~Mariooryad and C.~Busso, ``Correcting time-continuous emotional labels by
  modeling the reaction lag of evaluators,'' \emph{IEEE Transactions on
  Affective Computing}, vol.~6, no.~2, pp. 97--108, 2015, publisher: IEEE.

\bibitem{Huang2015a}
\BIBentryALTinterwordspacing
Z.~Huang, T.~Dang, N.~Cummins, B.~Stasak, P.~Le, V.~Sethu, and J.~Epps,
  ``\BIBforeignlanguage{en}{An {Investigation} of {Annotation} {Delay}
  {Compensation} and {Output}-{Associative} {Fusion} for {Multimodal}
  {Continuous} {Emotion} {Prediction}},'' in
  \emph{\BIBforeignlanguage{en}{Proceedings of the 5th {International}
  {Workshop} on {Audio}/{Visual} {Emotion} {Challenge}}}.\hskip 1em plus 0.5em
  minus 0.4em\relax Brisbane Australia: ACM, Oct. 2015, pp. 41--48. [Online].
  Available: \url{https://dl.acm.org/doi/10.1145/2808196.2811640}
\BIBentrySTDinterwordspacing

\bibitem{Khorram2019}
S.~Khorram, M.~McInnis, and E.~Mower~Provost, ``Jointly {Aligning} and
  {Predicting} {Continuous} {Emotion} {Annotations},'' \emph{IEEE Transactions
  on Affective Computing}, vol. 3045, no.~c, pp. 1--16, 2019, arXiv: 1907.03050
  Publisher: IEEE.

\bibitem{Nilsson1960}
K.~A. Hallgren, ``Computing {Inter}-{Rater} {Reliability} for {Observational}
  {Data}: {An} {Overview} and {Tutorial},'' \emph{Gff}, vol.~82, no.~2, pp.
  218--226, 2012.

\bibitem{Chao2018a}
\BIBentryALTinterwordspacing
Y.-W. Chao, S.~Vijayanarasimhan, B.~Seybold, D.~A. Ross, J.~Deng, and
  R.~Sukthankar, ``\BIBforeignlanguage{en}{Rethinking the {Faster} {R}-{CNN}
  {Architecture} for {Temporal} {Action} {Localization}},'' in
  \emph{\BIBforeignlanguage{en}{2018 {IEEE}/{CVF} {Conference} on {Computer}
  {Vision} and {Pattern} {Recognition}}}.\hskip 1em plus 0.5em minus
  0.4em\relax Salt Lake City, UT: IEEE, Jun. 2018, pp. 1130--1139. [Online].
  Available: \url{https://ieeexplore.ieee.org/document/8578222/}
\BIBentrySTDinterwordspacing

\bibitem{fan2021pytorchvideo}
H.~Fan, T.~Murrell, H.~Wang, K.~V. Alwala, Y.~Li, Y.~Li, B.~Xiong, N.~Ravi,
  M.~Li, H.~Yang, J.~Malik, R.~Girshick, M.~Feiszli, A.~Adcock, W.-Y. Lo, and
  C.~Feichtenhofer, ``{PyTorchVideo}: {A} deep learning library for video
  understanding,'' in \emph{Proceedings of the 29th {ACM} international
  conference on multimedia}, 2021.

\bibitem{Gemmeke2017}
J.~F. Gemmeke, D.~P.~W. Ellis, D.~Freedman, A.~Jansen, W.~Lawrence, R.~C.
  Moore, M.~Plakal, and M.~Ritter, ``Audio {Set}: {An} ontology and
  human-labeled dataset for audio events,'' in \emph{Proc. {IEEE} {ICASSP}
  2017}, New Orleans, LA, 2017.

\bibitem{tsai}
\BIBentryALTinterwordspacing
I.~Oguiza, ``tsai - {A} state-of-the-art deep learning library for time series
  and sequential data,'' 2022, tex.howpublished: Github. [Online]. Available:
  \url{https://github.com/timeseriesAI/tsai}
\BIBentrySTDinterwordspacing

\bibitem{Truong2011}
K.~P. Truong, R.~Poppe, I.~De~Kok, and D.~Heylen, ``A multimodal analysis of
  vocal and visual backchannels in spontaneous dialogs,'' \emph{Proceedings of
  the Annual Conference of the International Speech Communication Association,
  INTERSPEECH}, pp. 2973--2976, 2011, iSBN: 19909772 (ISSN).

\end{thebibliography}



\begin{IEEEbiography}[{\includegraphics[width=1in,height=1in,clip,keepaspectratio]{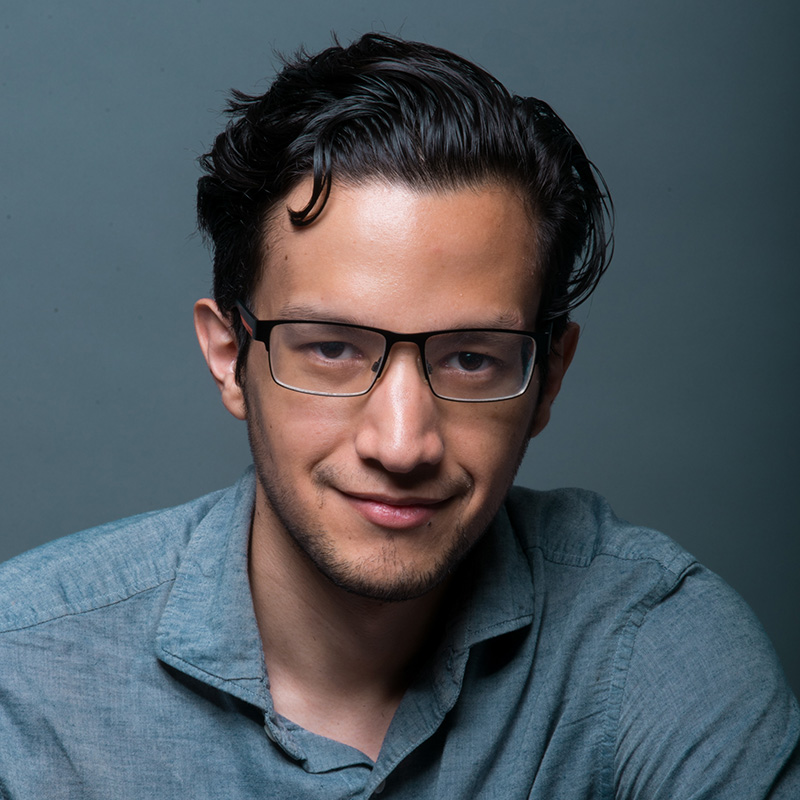}}]{Jose Vargas} is a PhD candidate at the Socially Perceptive Computing Lab at TU Delft, The Netherlands, since 2018. He is interested in multimodal action recognition and conversation quality assessment in-the-wild, the study of interpersonal adaptation and synchrony, and efficient annotation of in-the-wild data.
\end{IEEEbiography}

\begin{IEEEbiography}[{\includegraphics[width=1in,height=1in,clip,keepaspectratio]{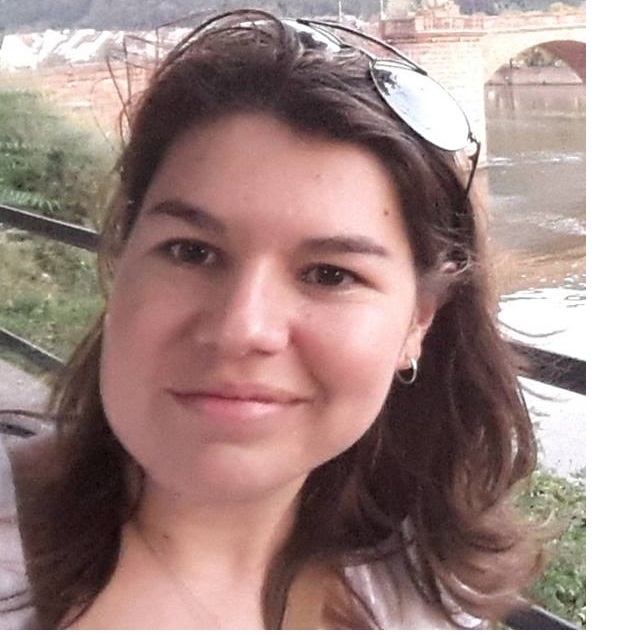}}]{Laura Cabrera-Quiros} is an Assistant Professor at the Costa Rican Institute of Technology (Instituto Tecnológico de Costa Rica), working in the Electronics Engineering department. Her research focuses on the use of machine learning and non-invasive technologies (e.g. wearable and embedded devices, cameras, physiological sensors) to understand human behavior, monitor health, and improve people’s quality of life.
\end{IEEEbiography}

\begin{IEEEbiography}[{\includegraphics[width=1in,height=1in,clip,keepaspectratio]{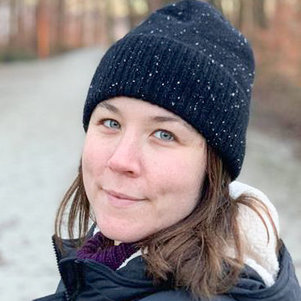}}]{Catharine Oertel} is an Assistant Professor at TU Delft, The Netherlands. She is co-principal investigator of the Designing Intelligence Lab (DI\_Lab), an effort aiming to bridge research done in computer science with industrial design engineering. Her research interest lies on understanding and modeling human interaction to build socially aware conversational agents able to engage with people in a human-like manner.
\end{IEEEbiography}

\begin{IEEEbiography}[{\includegraphics[width=1in,height=1in,clip,keepaspectratio]{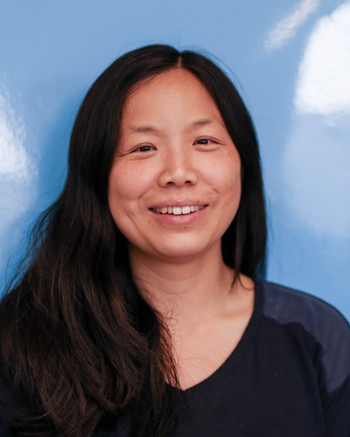}}]{Hayley Hung} is an Associate Professor in the Socially Perceptive Computing Lab at TU Delft, The Netherlands, where she works since 2013. Between 2010-2013 she held a Marie Curie Fellowship at the Intelligent Systems Lab at the University of Amsterdam. Between 2007-2010 she was a post-doctoral researcher at IDIAP Research Institute in Switzerland. She obtained her PhD in Computer Vision from Queen Mary University of London in 2007. Her research interests are social computing, social signal processing, computer vision, and machine learning.
\end{IEEEbiography}




\appendices

\section{Annotation experiment details}\label{app:annotation}

In this section we present some statistics from our annotation experiments from 48 annotators, including number of times the annotator detected laughter, distribution of intensity and confidence ratings, and time taken in completing the experiment. Additionally, we requested an optional rating of their experience completing the HIT (\textit{How would you rate your experience in completing this experiment?}) in a scale from 1 to 5.

Table \ref{tab:hit_details} shows the annotation details, with each row being one hit/annotator. Note that per our pre-annotations of laughter, 59 of the 84 examples in each HIT contained laughter, while the rest only contained speech. Note also that time taken in completing the experiment was measured as the difference between timestamps in the data sent at the beginning and end of the experiment, and does not contemplate the fact that annotators could have taken breaks in between.

We also allowed annotators to provide free text feedback (\textit{Do you have any comments about the process? Did you find it frustrating, tiring or too long? Were the instructions clear? Did you have any issues with the tool?}) about the process. Here, most annotators who answered reported surprise at the originality of the task, some saying they found it interesting and/or they had never completed an experiment of this kind. Some commented about the instructions, reporting them to be clear. Some annotators reported the experiment being long / tiring.

\begin{table*}
\centering
\caption{Details of the annotation HITs. \textit{G} indicates the HIT group. HITs within the same group contain the same laughter/non-laughter samples. \textit{N} indicates the HIT number within the group. HITs with the same \textit{N} are identical, except for the (random) ordering of the samples within each condition. HITs with different \textit{N} contain the same samples but assigned to different conditions. Each row corresponds to one annotator (HIT). \textit{\# positive} indicates the number of times laughter was detected by this person. \textit{Intensities} indicates the histogram of laughter intensities by each annotator (positive examples only). Each number (in order) corresponds to one step in the Likert scale (1-7).}
\label{tab:hit_details}
\begin{tabular}{rlllllrr}
\toprule
Annotator ID & G & N & \# positive &            Intensity &           Confidence & Time taken & Rating \\
\midrule
           1 & 0 & 1 &      53/84 & 13-11-11-09-06-03-00 & 03-09-07-07-08-15-35 &       43.0 &    4/5 \\
           2 & 0 & 1 &      55/84 & 17-19-09-05-04-02-00 & 02-02-08-06-06-13-47 &       45.5 &    4/5 \\
           3 & 0 & 2 &      49/84 & 10-11-12-04-08-02-02 & 03-04-09-03-15-16-27 &       41.6 &    4/5 \\
           4 & 0 & 2 &      61/84 & 08-10-10-01-12-08-00 & 00-00-01-01-08-15-54 &       41.9 &    5/5 \\
           5 & 0 & 3 &      56/84 & 03-10-21-15-02-03-01 & 03-00-01-44-24-11-01 &       93.8 &    3/5 \\
           6 & 0 & 3 &      63/84 & 13-16-13-08-11-05-01 & 02-09-05-05-16-24-23 &       33.4 &    5/5 \\
           7 & 1 & 1 &      55/84 & 03-09-09-13-16-05-00 & 00-02-07-04-10-13-48 &       40.7 &    5/5 \\
           8 & 1 & 1 &      59/84 & 06-17-14-06-09-06-02 & 00-01-02-02-10-20-49 &       36.5 &    5/5 \\
           9 & 1 & 2 &      48/84 & 15-13-08-06-10-07-03 & 04-06-05-01-08-13-47 &       47.8 &    4/5 \\
          10 & 1 & 2 &      57/84 & 14-06-11-02-14-06-03 & 00-00-06-03-11-13-51 &       48.8 &    5/5 \\
          11 & 1 & 3 &      63/84 & 22-17-16-13-10-06-00 & 18-25-18-09-10-04-00 &       54.2 &    5/5 \\
          12 & 1 & 3 &      57/84 & 10-11-13-16-05-07-00 & 00-01-01-24-07-09-42 &       47.8 &      - \\
          13 & 2 & 1 &      53/84 & 05-14-18-07-06-07-01 & 27-12-03-05-06-05-26 &       61.5 &    3/5 \\
          14 & 2 & 1 &      50/84 & 00-04-09-08-14-10-06 & 00-00-00-02-03-12-67 &       94.5 &    5/5 \\
          15 & 2 & 2 &      56/84 & 12-08-07-08-08-08-08 & 02-03-08-04-11-12-44 &       56.0 &    5/5 \\
          16 & 2 & 2 &      48/84 & 06-07-09-05-07-09-05 & 04-02-06-06-07-11-48 &       79.2 &    5/5 \\
          17 & 2 & 3 &      52/84 & 05-06-03-06-10-10-11 & 00-00-02-08-10-19-45 &       52.9 &    4/5 \\
          18 & 2 & 3 &      51/84 & 17-07-14-01-05-05-03 & 00-00-03-09-02-06-64 &       56.6 &    5/5 \\
          19 & 3 & 1 &      52/84 & 13-11-07-08-12-05-00 & 00-00-01-03-14-10-56 &       68.3 &    5/5 \\
          20 & 3 & 1 &      58/84 & 04-07-07-05-12-11-11 & 18-05-03-10-09-04-35 &       53.9 &    4/5 \\
          21 & 3 & 2 &      60/84 & 06-06-09-15-12-12-04 & 00-01-06-06-07-05-59 &       56.7 &    5/5 \\
          22 & 3 & 2 &      49/84 & 04-08-12-10-06-04-07 & 01-07-06-07-09-13-41 &       67.2 &    5/5 \\
          23 & 3 & 3 &      53/84 & 10-11-16-05-07-05-02 & 04-03-04-10-21-15-27 &       73.0 &    5/5 \\
          24 & 3 & 3 &      61/84 & 09-17-14-14-04-02-01 & 08-06-07-05-07-12-39 &       39.4 &    4/5 \\
          25 & 4 & 1 &      61/84 & 14-21-10-06-07-01-01 & 17-12-10-12-14-08-11 &       36.6 &    5/5 \\
          26 & 4 & 1 &      58/84 & 02-08-03-02-14-14-15 & 01-06-06-02-13-25-31 &       49.0 &    5/5 \\
          27 & 4 & 2 &      58/84 & 14-16-10-08-05-03-01 & 01-01-02-05-08-17-49 &       51.0 &    4/5 \\
          28 & 4 & 2 &      58/84 & 06-06-16-10-10-06-04 & 04-04-04-06-09-10-47 &       46.1 &    5/5 \\
          29 & 4 & 3 &      56/84 & 07-21-13-05-09-02-03 & 06-04-05-00-11-07-51 &       50.0 &    5/5 \\
          30 & 4 & 3 &      42/84 & 03-11-02-07-09-08-00 & 21-10-10-02-08-15-18 &       50.8 &    5/5 \\
          31 & 5 & 1 &      38/84 & 04-09-04-07-08-04-03 & 00-01-01-04-04-04-70 &      125.6 &    4/5 \\
          32 & 5 & 1 &      50/84 & 03-07-06-11-12-10-02 & 00-02-10-08-16-26-22 &       53.2 &    5/5 \\
          33 & 5 & 2 &      57/84 & 15-10-11-10-10-03-00 & 05-03-02-18-08-06-42 &       55.5 &    5/5 \\
          34 & 5 & 2 &      56/84 & 09-18-14-11-05-01-00 & 01-03-07-05-07-10-51 &       51.0 &    5/5 \\
          35 & 5 & 3 &      44/84 & 07-07-08-11-05-02-00 & 00-00-02-09-41-15-17 &       73.2 &    4/5 \\
          36 & 5 & 3 &      56/84 & 11-10-08-11-09-09-08 & 03-10-08-07-13-22-21 &       60.3 &    5/5 \\
          37 & 6 & 1 &      58/84 & 08-14-12-11-08-04-00 & 00-07-13-06-28-21-09 &       53.8 &    4/5 \\
          38 & 6 & 1 &      51/84 & 16-19-15-04-02-00-00 & 01-05-07-09-12-20-30 &       72.2 &    4/5 \\
          39 & 6 & 2 &      53/84 & 10-06-10-05-14-07-03 & 00-00-02-05-08-15-54 &       42.0 &      - \\
          40 & 6 & 2 &      46/84 & 01-06-06-03-15-11-04 & 04-06-10-02-14-06-42 &       54.6 &    5/5 \\
          41 & 6 & 3 &      51/84 & 14-09-08-07-08-07-01 & 05-06-01-03-06-02-61 &       35.5 &      - \\
          42 & 6 & 3 &      48/84 & 07-12-07-10-09-04-00 & 02-05-09-03-17-13-35 &       40.0 &    5/5 \\
          43 & 7 & 1 &      38/71 & 14-14-07-07-03-00-00 & 03-05-03-04-04-15-37 &       39.7 &    4/5 \\
          44 & 7 & 1 &      40/71 & 01-07-09-05-07-05-05 & 03-05-10-06-15-18-14 &       32.2 &    5/5 \\
          45 & 7 & 2 &      47/71 & 06-08-09-13-13-02-00 & 01-07-08-05-16-32-02 &       31.3 &    5/5 \\
          46 & 7 & 2 &      55/71 & 12-10-13-08-10-06-00 & 03-04-10-01-23-10-20 &       60.3 &    4/5 \\
          47 & 7 & 3 &      45/71 & 05-20-11-04-05-00-00 & 00-06-05-02-04-11-43 &       44.9 &    5/5 \\
          48 & 7 & 3 &      46/71 & 08-13-14-05-04-02-00 & 03-03-05-02-05-19-34 &       33.8 &    5/5 \\
\bottomrule
\end{tabular}
\end{table*}

\section{Additional Experiments}

\begin{figure}
    \centering
    \includegraphics[width=\columnwidth]{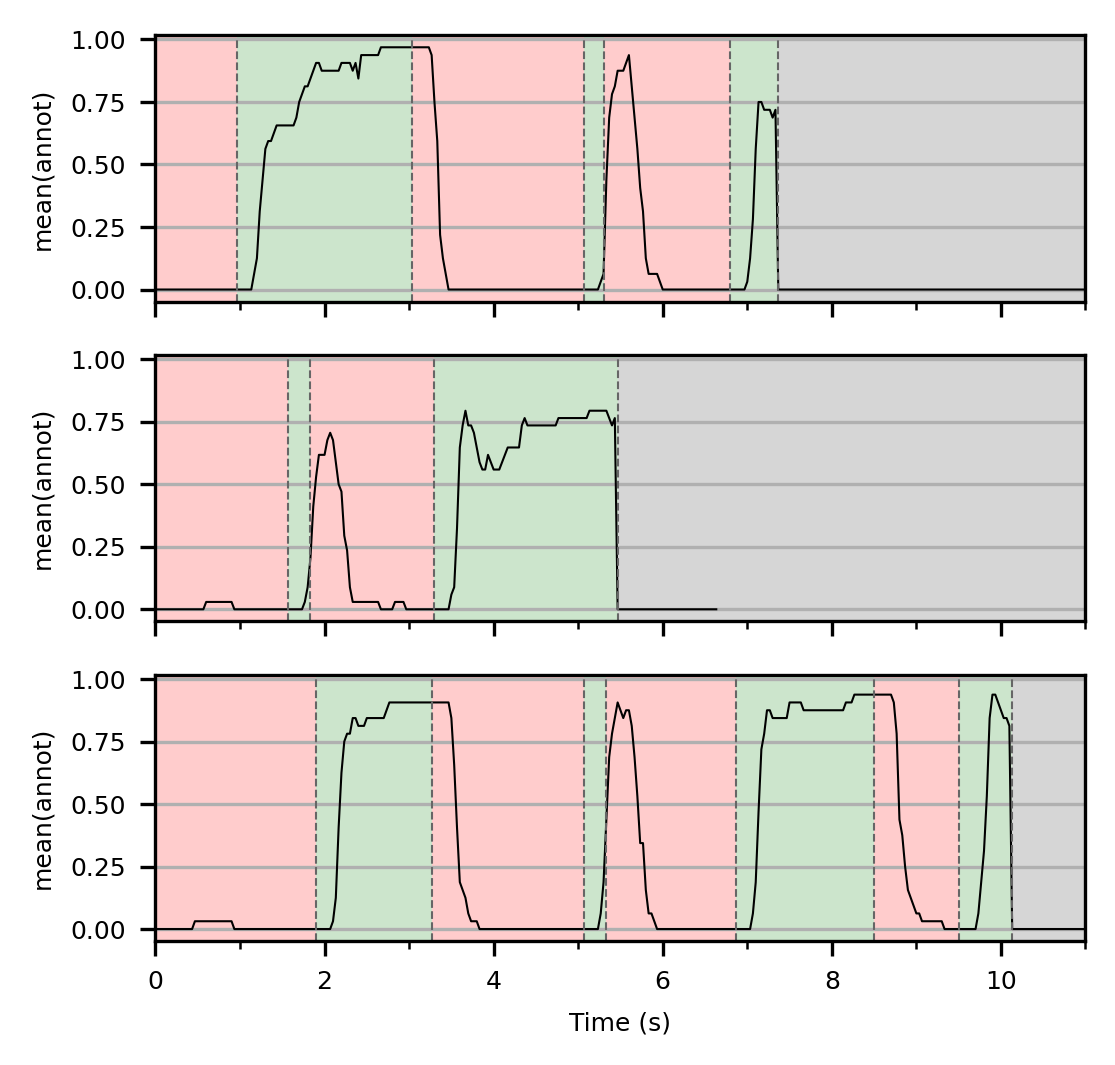}
    \caption{Mean value of the annotations during the reaction time experiments. Grayed out intervals indicate when the circle was present in the screen. For these well-defined stimuli the measured delays are more similar (steep slopes).}
    \label{fig:actions:reaction}
\end{figure}

In an attempt to obtain a better measure of reaction time to simpler stimuli, each annotator also completed three tasks designed for this goal. These tasks consisted in short videos (no audio), in which a circle would appear in the center at random intervals. Annotators were instructed to press a key to indicate when the circle was present in the video, as when annotating actions. Figure \ref{fig:actions:reaction} plots the mean value of the annotations (across annotators) for the three reaction time tasks.

The steeper onsets and offsets in these idealistic experiments serve as validation that the curves observed in figure \ref{fig:actions:onsets} are due to differences in perception of the phenomenon itself, rather than reaction time alone. This is likely due to annotators having different mental models of the actions being performed, at least when it comes to their dynamics.

To incorporate this notion into our analysis, we computed IoUs for different values of delay, where the delay value is used to shift the continuous annotations before computation of the IoU.

\section{Method details}\label{app:methods}

\subsection{Segmentation methods}

\section{HIT Structure}\label{app:hit_struct}

The HITs in our human annotation study consisted in a sequence of pages or tasks to be followed by an annotator in order. In general, each HIT contained several introductory tasks and examples, followed by three annotation blocks, one for each modality condition. The order of these three blocks, and of the examples within was randomized for each instance.

In detail, each HIT consists of the following tasks:

\begin{enumerate}
    \item Consent Form (5 min). Participants were asked to agree to an End User License Agreement required to access our dataset. This was put in place to protect the privacy of data subjects and was required to continue with the annotation.
    
    \item General instructions (5 min). Introduction to the HIT, informing the annotator about the different sections/conditions, the need for audio equipment and the structure of the HIT.
    
    \item Reaction time test explainer (2 min). An example reaction time test, with instructions to let the annotator familiarize themselves with the test.
    
    \item Example laughter segments (3 min). Three example segments where the annotator was asked to continuously annotate and then rate laughter segments. We chose segments were laughter was clear and evident since the sole purpose of these segments was to let the subjects become familiar with the process.
    
    \item Video-only block (28 segments, 10 min). One video per page. The participant must play the video and press a keyboard key when they perceive laughter to be occurring. At the end of each segment the participant must provide a rating of laughter intensity using a slider with a continuous range between 0 and 10 and a rating of confidence in their laughter annotation. This is all explained in a page with instructions at the start of the block.
    
    \item Audio-only block (28 segments, 10 min). Same as above. Participants must play a web video containing no image (only audio) and similarly press a key when they think they can hear laughter. In the instructions page, participants are asked to test their audio equipment (speakers, headphones).
    
    \item Audiovisual block (28 segments, 10 min). Same as above, but now annotators get access to both audio and video.
    
    \item Optional feedback (1 min). Annotators were asked to (optionally) rate their experience and give free text feedback on the process.
\end{enumerate}

\end{document}